\documentclass[twocolumn]{aastex63} 
\usepackage[utf8]{inputenc}
\usepackage{rotating}
\usepackage{graphicx}

\usepackage{lineno}

\submitjournal{PSJ}
\accepted{}

\shorttitle{Structure and color of Jupiter’s clouds}
\shortauthors{Dahl et al.}

\begin{document}

\title{Vertical Structure and Color of Jovian Latitudinal Cloud Bands during the Juno Era}

\correspondingauthor{Emma Dahl}
\email{dahlek@nmsu.edu}

\author[0000-0003-2985-1514]{Emma K. Dahl}
\affiliation{New Mexico State University \\
PO Box 30001, MSC 4500 \\
Las Cruces, NM 88001, USA}

\author[0000-0002-9984-4670]{Nancy J. Chanover}
\affiliation{New Mexico State University \\
PO Box 30001, MSC 4500 \\
Las Cruces, NM 88001, USA}

\author[0000-0001-7871-2823]{Glenn S. Orton}
\affiliation{ Jet Propulsion Laboratory/California Institute of Technology \\
4800 Oak Grove Dr \\
Pasadena, CA 91109, USA}

\author[0000-0002-5544-9986]{Kevin H. Baines}
\affiliation{ Jet Propulsion Laboratory/California Institute of Technology \\
4800 Oak Grove Dr \\
Pasadena, CA 91109, USA}
\affiliation{Space Science and Engineering Center\\
University of Wisconsin–Madison\\
Madison, WI 53706, USA}

\author[0000-0001-5374-4028]{James A. Sinclair}
\affiliation{ Jet Propulsion Laboratory/California Institute of Technology \\
4800 Oak Grove Dr \\
Pasadena, CA 91109, USA}

\author{David G. Voelz}
\affiliation{New Mexico State University \\
PO Box 30001, MSC 4500 \\
Las Cruces, NM 88001, USA}

\author[0000-0002-0766-9802]{Erandi A. Wijerathna}
\affiliation{New Mexico State University \\
PO Box 30001, MSC 4500 \\
Las Cruces, NM 88001, USA}

\author{Paul D. Strycker}
\affiliation{Concordia University Wisconsin \\
12800 N Lake Shore Drive\\
Mequon, WI 53097, USA}

\author[0000-0002-6772-384X]{Patrick G. J. Irwin}
\affiliation{Atmospheric, Oceanic \& Planetary Physics\\
University of Oxford\\
Parks Road, Oxford, OX1 3PU, UK}

\begin{abstract}

The identity of the coloring agent(s) in Jupiter's atmosphere and the exact structure of Jupiter's uppermost cloud deck are yet to be conclusively understood. The Cr\`{e}me Br\^ul\'ee model of Jupiter's tropospheric clouds, originally proposed by \citet{baines_2014} and expanded upon by \citet{sromovsky_2017} and \citet{baines_2019}, presumes that the chromophore measured by \citet{carlson_2016} is the singular coloring agent in Jupiter's troposphere. In this work, we test the validity of the Cr\`{e}me Br\^ul\'ee model of Jupiter's uppermost cloud deck using spectra measured during the \textit{Juno} spacecraft's 5$^{\mathrm{th}}$ perijove pass in March 2017. These data were obtained as part of an international ground-based observing campaign in support of the \textit{Juno} mission using the NMSU Acousto-optic Imaging Camera (NAIC) at the 3.5-m telescope at Apache Point Observatory in Sunspot, NM. We find that the Cr\`{e}me Br\^ul\'ee model cloud layering scheme can reproduce Jupiter's visible spectrum both with the \citet{carlson_2016} chromophore and with modifications to its imaginary index of refraction spectrum. While the Cr\`{e}me Br\^ul\'ee model provides reasonable results for regions of Jupiter's cloud bands such as the North Equatorial Belt and Equatorial Zone, we find that it is not a safe assumption for unique weather events, such as the 2016-2017 Southern Equatorial Belt outbreak that was captured by our measurements.

\end{abstract}


\section{Introduction} \label{sec:intro}

Jupiter's atmosphere is a dynamic, variable, and complicated system \citep{jupbook2,jupbook_dynamics}. The measurements made by the \textit{Juno} spacecraft since its arrival at Jupiter in July 2016 have wholly recontextualized the Jovian atmosphere, magnetosphere, interior, and system as a whole \citep{bolton_2017,bolton_2019}. However, even with the detailed measurements and unprecedented resolution afforded by \textit{Juno} and other spacecraft, as well as decades of ground-based measurements, a consistent picture of the vertical structure of Jupiter's colorful clouds, the source of those colors, and the mechanisms driving variations in color and storm patterns observed at the cloud tops remains elusive.

\begin{figure*}
\centering
\includegraphics[scale=0.6]{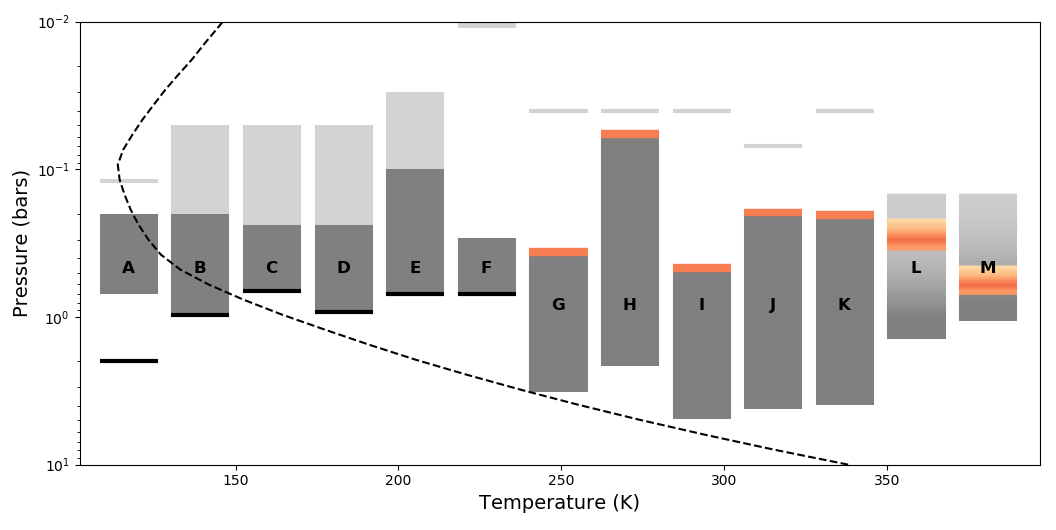}
\caption{Figure updated from \citet{sromovsky_2002} illustrating the variety of cloud base pressures and structures that can be used to reproduce spectra of various atmospheric regions on Jupiter. The parameters and corresponding references describing these results are in Table \ref{tab:atm_structure_table}. Black lines indicate opaque tropospheric sheet clouds; dark gray blocks indicate either tropospheric hazes or clouds; light gray lines and blocks indicate stratospheric hazes; orange bars indicate chromophore layers/regions (Models A-C include chromophore elements but not in the same discrete layer as Models D-J); gray and orange gradients represent aerosol and chromophore profiles respectively that contain clouds and/or hazes that vary much more with altitude than the simple clouds of models A-H.}
\label{fig:atm_structure}
\end{figure*}

The location and composition of Jupiter’s tropospheric cloud layers, while difficult to measure directly, have been predicted through the use of thermochemical equilibrium models. These models use temperature-pressure profiles, chemical abundances, chemical reaction paths, and temperature- and pressure-dependent reaction rates to predict the depths at which certain species will condense. Water, ammonium hydrosulfide, and ammonia have been predicted to condense into Jovian clouds at approximately 6, 2.2, and 0.7 bars, respectively \citep{lewis_1969,Weidenschilling_1973,atreya_1999}. 

While the locations of these predicted cloud bases have long been taken as the starting point for interpreting observations of Jupiter's atmosphere, ground- and space-based observations and modeling results point to the likelihood that these base pressures are not safe assumptions \citep{jupbook2}. Most of the observed contrast variation in Jupiter’s uppermost cloud deck can be explained by models containing (but not limited to) a stratospheric haze layer above a main cloud deck. This main cloud deck is often either presumed to have a base somewhere within the 0.4-1.4 bar range \citep{chanover_1997, banfield_1998, irwin_1998, sromovsky_2002, strycker_2011, perez_hoyos_2012, giles_2015, braude_2020}, or it is modeled at deeper pressures, from 2-4 bars or below \citep{smith_1986,baines_2019,sromovsky_2017}. While such models have been able to account for most observations, discrepancies remain between model results for different datasets with different spectral and spatial resolutions, even when those datasets lie in the same spectral regime \citep{sromovsky_2002}. This problem is exacerbated by both physical changes in Jupiter's atmosphere over time and major parameter degeneracies in the visible and near-infrared wavelength regimes. Figure \ref{fig:atm_structure} and Table \ref{tab:atm_structure_table} show a summary of results from multiple studies of the vertical structure of several different major Jovian cloud regions in order to illustrate these discrepancies. In Table \ref{tab:atm_structure_table}, $\tau$ is the optical depth of a given layer, $r$ the particle size in microns, and $P_{base}$ and $P_{top}$ are the pressures at the bottom and top of the layer, respectively; parameters denoted with * indicate values that were held fixed for a given analysis.

\begin{longrotatetable}
\begin{deluxetable*}{ccclcccc}
\tablecaption{Summary of cloud structure parameters corresponding to Fig. \ref{fig:atm_structure} \label{tab:atm_structure_table}}
\tablewidth{700pt}
\tabletypesize{\scriptsize}
\tablehead{
\colhead{Model} & \colhead{Source} & 
\colhead{Cloud region} & \colhead{Model layers} & 
\colhead{$\tau$} & \colhead{$r$ $(\mu m)$} & 
\colhead{$P_{base}$ (bar)} & \colhead{$P_{top}$ (bar)}
} 
\startdata
    A & \cite{smith_1986} & SEB & 1 - Tropospheric sheet cloud & 8-24 & 0.63-0.95 & 2.0 & 2.0 \\
      & & & 2 - Tropospheric haze & 3-4 & 0.63-0.95 & 0.7 & 0.13 \\
      & & & 3 - Stratospheric haze & 0.3 & 0.64 & 0.1 & 0.03 \\
    \hline
    B & \citet{simon_2001b} & NEB & 1 - Tropospheric sheet cloud & 5.5$\pm$1.1-1.65 &  0.9-3.0 & 0.97$\pm$0.145 & 0.97$\pm$0.145 \\
    & & & 2 - Tropospheric haze & 6$\pm$1.2-1.8 & 0.6-1.2 & 0.97$\pm$0.145 & 0.2$\pm$0.03 \\
    & & & 3 - Stratospheric haze & 0.03$\pm$0.006-0.009 & 0.01-0.05 & 0.2$\pm$0.03 & - \\ 
    \hline 
    C & \cite{simon_2001b} & EZ & 1 - Tropospheric sheet cloud & 38$\pm$7.6-11.4 & 0.9-3.0 & 0.67$\pm$0.1 & 0.67$\pm$0.1 \\
    & & & 2 - Tropospheric haze & 4$\pm$0.8-1.2 & 0.6-1.2 & 0.67$\pm$0.1 & 0.24$\pm$0.036 \\
    & & & 3 - Stratospheric haze & 0.2$\pm$0.04-0.06 & 0.01-0.05 & 0.24$\pm$0.036 & - \\
    \hline
    D & \cite{simon_2001b} & SEB & 1 - Tropospheric sheet cloud & 5.6$\pm$1.12-1.68 & 0.9-3.0 & 0.93$\pm$0.14 & 0.93$\pm$0.14 \\
    & & & 2 - Tropospheric haze & 9.3$\pm$1.86-2.79 & 0.6-1.2 & 0.93$\pm$0.14 & 0.24$\pm$0.036 \\
    & & & 3 - Stratospheric haze & 0.09$\pm$0.018-0.027 & 0.01-0.05 & 0.24$\pm$0.036 & - \\
    \hline
    E & \cite{strycker_2011} & EZ & 1 - Tropospheric sheet cloud & 29.0$\pm$ 6.0* & 2.0* & (0.4, 0.7, 1.0)* & (0.4, 0.7, 1.0)* \\
      & & & 2 - Tropospheric haze & 3.5 $\pm$ 0.4 & 0.9* & (0.4, 0.7, 1.0)* & 0.2* \\
      & & & 3 - Stratospheric haze & 0.22 $\pm$ 0.02 & 0.03* & 0.1* & 0.1* \\
      \hline
   F & \cite{perez_hoyoz_2012_SEB} & SEB (Fade) & 1 - Tropospheric sheet cloud & 100$\pm$50 & - & 0.7* & 0.7* \\
   & & & 2 - Tropospheric haze & 4.4$\pm$0.4 & 1.0$\pm$0.5 & - & 0.29$\pm$0.03\\
   & & & 3 - Stratospheric haze & <0.15 & 0.3$\pm$0.1 & 0.1* & 0.001* \\
      \hline
    G & \cite{sromovsky_2017} & NEB & 1 - Tropospheric cloud & $16.061_{-0.995}^{+1.057}$ & $1.43_{-0.213}^{+0.220}$ & $3.213_{-0.230}^{+0.243}$ & 0.381 $\pm$ 0.017 \\
      & & & 2 - Chromophore layer & $0.186_{-0.021}^{+0.023}$ & $0.151_{-0.009}^{+0.010}$ & 0.381 $\pm $0.017 & 0.342 $\pm$ 0.015 \\
      & & & 3 - Stratospheric haze & $0.01_{-0.003}^{+0.004} $ & 0.1* & 0.04* & 0.04* \\
      \hline
    H & \cite{sromovsky_2017} & EZ & 1 - Tropospheric cloud & $13.663_{-1.232}^{+1.348}$ & $0.586_{-0.059}^{+0.066}$ & $2.15_{+0.151}^{-0.138}$ & $0.06_{-0.011}^{+0.034}$ \\
      & & & 2 - Chromophore layer & $0.059_{-0.010}^{+0.012}$ & $0.117_{-0.012}^{+0.014}$ & $0.06_{-0.011}^{+0.034}$ & $0.054_{-0.009}^{+0.03}$ \\
      & & & 3 - Stratospheric haze & 0.0 $\pm$ 0.0 & 0.1* & 0.04* & 0.04* \\
      \hline
    I & \cite{sromovsky_2017} & SEB & 1 - Tropospheric cloud & $25.187_{-3.631}^{+4.179}$ & $0.836_{-0.193}^{+0.239}$ & $4.9_{-0.741}^{+0.703}$ & 0.489 $\pm$ 0.018 \\
      & & & 2 - Chromophore layer & $0.757_{-0.064}^{+0.058}$ & $0.286_{-0.014}^{+0.015}$ & 0.489 $\pm$ 0.018 & 0.449 $\pm$ 0.016 \\
      & & & 3 - Stratospheric haze & 0.027 $\pm$ 0.003 & 0.1* & 0.04* & 0.04* \\
      \hline
    J & \cite{sromovsky_2017} & GRS & 1 - Tropospheric cloud & $29.56_{-3.76}^{+4.24}$ & $1.148_{-0.245}^{+0.277}$ & $4.192_{-0.614}^{+0.630}$ & 0.205 $\pm$ 0.012 \\
      & & & 2 - Chromophore layer & $0.209_{-0.020}^{+0.021}$ & 0.149 $\pm$ 0.007 & 0.205 $\pm$ 0.012 & 0.18 $\pm$ 0.010 \\
      & & & 3 - Stratospheric haze & $0.004_{-0.002}^{+0.004}$ & 0.1* & 0.07* & 0.07* \\
      \hline
    K & \cite{baines_2019} & GRS & 1 - Tropospheric cloud & $28.5_{-4.8}^{+5.3}$ & $1.08_{-0.31}^{+0.38}$ & $3.94_{-0.68}^{+0.86}$ & $0.212_{-0.019}^{+0.022}$ \\
      & & & 2 - Chromophore layer & $0.194_{-0.026}^{+0.018}$ & $0.141_{-0.011}^{+0.013}$ & $0.212_{-0.019}^{+0.022}$ & $0.19_{-0.017}^{+0.019}$ \\
      & & & 3 - Stratospheric haze & $0.074_{-0.35}^{+0.41}$ & 0.25* & 0.04* & 0.04* \\
      \hline
    L & \cite{braude_2020} & EZ & 1 - Tropospheric cloud and haze & 25.52$\pm$2.67 & 4.40 $\pm$ 0.05 & 1.4 $\pm$ 0.1** & 0.15* \\
      & & & 2 - Chromophore region & 0.0306$\pm$0.017 & 0.05* & 0.3$\pm$ 0.02** & - \\
      \hline
    M & \cite{braude_2020} & NEB & 1 - Tropospheric cloud and haze & 19.97$\pm$2.85 & 1.5 $\pm$ 0.2 & 1.07 $\pm$ 0.08** & 0.15* \\
      & & & 2 - Chromophore region & 0.0687$\pm$0.02 & 0.05* & 0.61 $\pm$ 0.04** & - \\
\enddata
\tablecomments{Parameters denoted with * indicate values that were held fixed for a given analysis. ** denotes the level of maximum optical depth for a given model layer.}
\end{deluxetable*}
\end{longrotatetable}

Besides the structure of the cloud decks, the chemical composition, number of, and horizontal and vertical distribution of the chromophores that give Jupiter's bands and storms their distinct reddish hue is also an ongoing area of study. \citet{simon_2001a} used limited Hubble Space Telescope data and the method of principal component analysis (PCA) to identify three components contributing to Jupiter's brightness variations. They found that overall brightness differences aside from color accounted for $\sim$91\% of variation within the image, a blue absorber was responsible for $\sim$8\% of variations in or around the tropospheric cloud deck, and a second coloring agent was necessary to explain the remaining $\sim$1\% of variation in anticyclonic systems such as the Great Red Spot. After comparing features between different datasets, \citet{simon_2001a} notes the possibility of either a white cloud deck covered with a uniform layer of some blue absorber or of overall pink clouds (although it should be remembered that these datasets were spectrally limited). A study completed with \textit{Galileo} Solid State Imager data showed that the main color difference between the dark North Equatorial Belt and bright Equatorial Zone lay in the 410-nm single scattering albedo in a tropospheric haze layer, and that such color differences were not seen at redder wavelengths \citep{simon_2001b}. Additionally, \citet{ordonez_etxeberria_2016} used several analysis techniques including PCA to examine the color of Jupiter's clouds and hazes as seen by \textit{Cassini}'s Imaging Science Subsystem. This study also found that a single chromophore could explain the color variations throughout Jupiter's atmosphere, with the exception of some small reddish-brown cyclones within the North Equatorial Belt, which needed two coloring agents.

Recent laboratory investigations into the chemical identity of the chromophore(s) include a study of irradiated ammonium hydrosulfide \citep{loeffler_2016, loeffler_2018} and an analysis of photolyzed ammonia reacting with acetylene \citep{carlson_2016}. Contemporary modeling work using the chromophore discussed in \citet{carlson_2016} showed that this ammonia-based molecule, when present in a relatively thin layer directly above the uppermost cloud deck (in addition to a separate stratospheric haze layer), can effectively reproduce spectra of Jupiter's atmosphere as observed by the Visual and Infrared Mapping Spectrometer (VIMS) instrument onboard the \textit{Cassini} spacecraft during its flyby of Jupiter in late December, 2000 \citep{baines_2014,baines_2016,baines_2019, sromovsky_2017}. This cloud layering scheme, which has been dubbed the Cr\`{e}me Br\^ul\'ee model (which will be denoted as the CB model throughout the rest of this analysis), can reproduce varying degrees of redness by varying the opacity of the thin chromophore layer, making the \citet{carlson_2016} chromophore universal throughout the troposphere. These analyses also tested variations of the CB model, including allowing the chromophore layer to diffuse upward from the main cloud \citep{sromovsky_2017}, placing the chromophores into a stratospheric haze \citep{baines_2016,baines_2019}, and coating cloud particles with a chromophore material \citep{baines_2019}. However, the CB model consistently produced better fits to spectra. The results from \citet{sromovsky_2017} and \citet{baines_2019} show that an ammonia-dominated uppermost cloud deck with a thin chromophore layer directly above it can provide high-quality fits to VIMS data with main cloud bases in the $\sim$2-5 bar range. 

While \citet{sromovsky_2017} and \citet{baines_2019} could reproduce Jupiter's visible spectrum with the CB model, \citet{braude_2020} showed that both the cloud layering scheme of the CB model and the assumption that the \citet{carlson_2016} chromophore was the universal coloring agent within the chromophore layer were unable to successfully reproduce spectra of the Great Red Spot and other major banded regions of Jupiter's atmosphere. The datasets analyzed in \citet{braude_2020} were much more recent than those analyzed in  \citet{sromovsky_2017} and \citet{baines_2019} and were obtained in the pre-\textit{Juno} era and in conjunction with the \textit{Juno} spacecraft's 6$^{\mathrm{th}}$ and 12$^{\mathrm{th}}$ perijove passes with the Very Large Telescope/Multi Unit Spectroscopic Explorer (VLT/MUSE) instrument. To rectify the CB model's inability to fit their spectra, they retrieved a new chromophore from spectra of the North Equatorial Belt that could replace the \citet{carlson_2016} coloring agent. They then applied this new universal chromophore to a continuous cloud and haze profile (as opposed to the sheet clouds of the CB model) that placed the base of the main cloud at $\sim$1.0-1.4 bars. 

These bodies of work all tested the validity of the CB model, but with varying results: \citet{sromovsky_2017} and \citet{baines_2019} successfully reproduced visible and near-infrared spectra of Jupiter's atmosphere using this model, but \citet{braude_2020} found that the CB model could not reproduce their data. In this analysis, with hyperspectral image cubes observed with the Astrophysical Research Consortium 3.5-m telescope at Apache Point Observatory (APO) in Sunspot, NM, we aim to also test the validity the CB model as a parameterization of Jupiter's tropospheric clouds.

The data we describe in the next section were obtained as part of an international campaign of ground-based observers that were mobilized to acquire observations of Jupiter in complementary wavelength regimes to those observed by the \textit{Juno} spacecraft\footnote{https://www.missionjuno.swri.edu/planned-observations} \citep{junocam,orton_2017}. \textit{Juno} has two instruments that have been measuring the Jovian atmosphere at depth in the infrared and microwave regimes (the Jovian InfraRed Auroral Mapper and the Microwave Radiometer,  respectively), as well as JunoCam, a visible camera being used to image the cloud tops. JunoCam’s four broadband visible filters cover the red, green, and blue parts of the visible spectrum as well as a methane absorption band, with center wavelengths of 698.9, 553.5, 480.1, and 893.3 nm, respectively \citep{junocam}. Other than the relatively narrow methane filter, JunoCam’s RGB filters have an average full width at half maximum (FWHM) of $\sim$100 nm. While the suite of instruments on board \textit{Juno} allows for unprecedented spatial resolution and detailed measurements of Jupiter’s atmosphere at depth, the wide bandwidths of JunoCam’s visible filters prohibit detailed sampling of Jupiter’s cloud tops. Therefore, visible hyperspectral image cubes acquired during \textit{Juno}'s close perijove passes are a highly complementary dataset to measurements made by \textit{Juno}, providing context for measurements of the deeper atmosphere and expanding \textit{Juno}'s resolution from narrow longitudinal swaths to disk-wide coverage.

In this study, we present the results of radiative transfer modeling and analysis of visible spectra of Jupiter's Equatorial Zone, North Equatorial Belt, and an outbreak cloud in the South Equatorial Belt as measured during \textit{Juno}’s 5$^{\mathrm{th}}$ perijove (PJ) pass in March 2017. We use the CB model of Jupiter's atmosphere to parameterize Jupiter's uppermost cloud deck and seek to test both the universality of the \citet{carlson_2016} chromophore and the ability of the CB cloud layering scheme to reproduce our spectra. In Section 2, we discuss the instrument used to collect our image cubes, the data reduction and calibration process, and the resulting data products. In Section 3, we review the radiative transfer code used to model these data and our modeling methodology. In Section 4, we explore the best fit atmospheric models and summarize the results. In Section 5, we discuss these results and their limitations, speculate on the physical mechanisms that might cause them, review differences and similarities between this and previous work, and briefly compare them to observations made by \textit{Juno}.

\section{Observations} \label{sec:obs}

As part of an international ground-based observing campaign in support of the \textit{Juno} mission \citep{junocam,orton_2017}, we used the New Mexico State University (NMSU) Acousto-optic Imaging Camera (NAIC) to obtain hyperspectral image cubes of Jupiter in the visible wavelength regime during \textit{Juno} perijove passes whenever viewing geometry allowed. We used these data to produce I/F spectra within three major cloud bands on Jupiter: the Equatorial Zone (EZ), North Equatorial Belt (NEB), and an outbreak within the South Equatorial Belt (SEB). Due to viewing geometry limitations unique to our image cube acquisition strategy and the scope of our science objectives, we limit our analysis to locations near the sub-observer longitude of the planet.

\subsection{Instrument description}

NAIC is a hyperspectral imager that uses a charge-coupled device (CCD) camera to record images and an acousto-optic tunable filter (AOTF) as its filtering element. AOTFs are optical filters that take advantage of the diffractive qualities of birefringent crystals with high elasto-optic coefficients ({\it i.e.}, crystals whose index of refraction changes readily in the presence of an acoustic wave). When a radio-frequency acoustic wave is applied to such crystals, they behave like a phase grating, thus allowing the user to ``tune" the crystal to produce a filter centered on the wavelength of choice \citep{glenar}. After incident broadband light is diffracted through the crystal, four total beams emerge from the AOTF: two diffracted narrowband beams at equal and opposite angles of diffraction, and two rays of undiffracted broadband light. One or both of the diffracted, narrowband, ``tuned" beams can then be imaged on a detector. Owing to their low masses, dynamic spectral tuning abilities, and lack of moving parts, AOTFs have been flown on several spacecraft, including Venus Express and Hayabusa-2 \citep{korablev_2018}. For our purposes, AOTFs are especially well-suited for sampling Jupiter's atmosphere with numerous narrow filters, providing us with the pressure sampling necessary to distinguish differences in cloud structure.

\begin{figure*}
\centering
\includegraphics[scale=0.6]{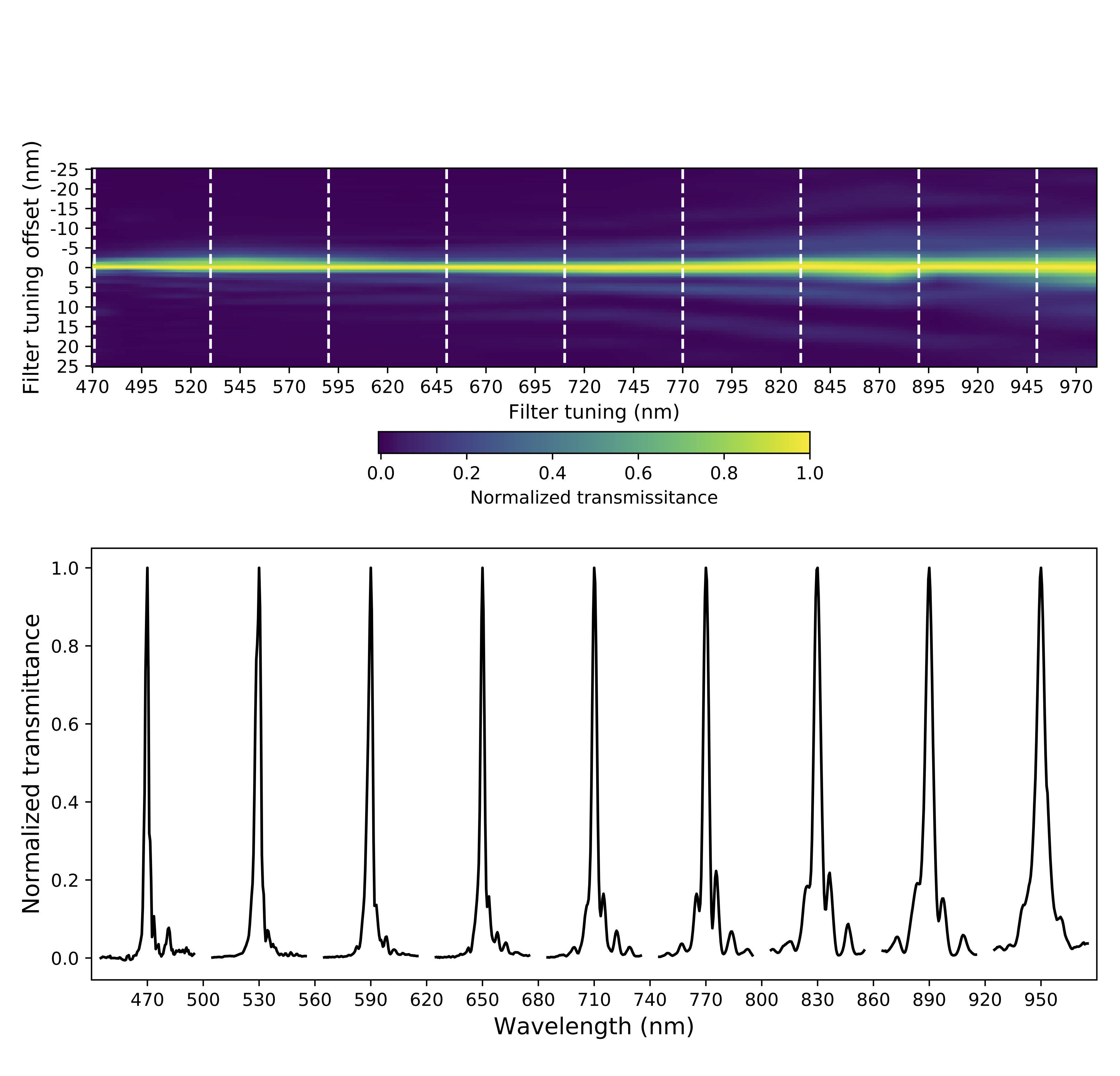}
\caption{Top: Normalized NAIC filter transmission as a function of the filters' central (tuned) wavelength and the offset from the center wavelength. 
Bottom: 9 selected filter functions at 470, 530, 590, 650, 710, 770, 830, 890, and 950 nm, corresponding to the dashed white lines in the above panel. NAIC uses 241 such filters in total when acquiring a hyperspectral image cube.}
\label{fig:filters}
\end{figure*}

When observing, we operated our AOTF from 470 to 950 nm and took an image every 2 nm, producing an average spectral resolution over this wavelength range ($R = \lambda/ \Delta \lambda$) of 205.7. The filter transmission functions of NAIC resemble \textit{sinc\textsuperscript{2}} functions (where $sinc^2(x) = 1$ for $x=0$ and $= sin^2(x)/x^2$ otherwise) whose central and side lobes widen with wavelength; these filter transmission functions and their variation with wavelength are illustrated in Figure \ref{fig:filters}. For the observations described in this analysis, we used NAIC at the Astrophysical Research Consortium 3.5-m telescope at Apache Point Observatory (APO) in Sunspot, NM. Our diffraction-limited plate scale for these observations was 0.1027 $"$/pixel. Specifications for the AOTF and detector as used on the telescope are in Table \ref{tab:instrument_specs}.

\begin{deluxetable}{ c c }
\tablecaption{NAIC setup and specifications while mounted to the 3.5-m telescope at APO\label{tab:instrument_specs}}
\tablehead{\colhead{Characteristic} & \colhead{Value}}
\tablewidth{0pt}
\startdata
    \multicolumn{2}{c}{\textbf{AOTF}}  \\  
    Crystal tunable range & 450-1000 nm \\
    Average FWHM* & 3.54 nm \\
    Average $\lambda/ \Delta \lambda$* & 205.7 \\
    \multicolumn{2}{c}{\textbf{Detector}}  \\ 
    CCD model & Apogee Alta F47 \\
    CCD chip size (unbinned) & 1024$\times$1024 pixels\\
    Pixel size & 13 $\mu m$\\
    Plate scale (binned 2$\times$2) & 0.1027$"$/pixel \\
    FOV & 52.58$"$ $\times$52.58$"$ \\
\enddata
\tablecomments{*Averages were taken since in general, as the wavelength of the filter increases so does the full width at half maximum (FWHM) of its central lobe. Stated averages are over the range of wavelengths used in an image cube (470-950 nm), not over the entire tuning range of the crystal}
\end{deluxetable}

\subsection{Data Selection and Calibration}

The data analyzed for this work were obtained during \textit{Juno}'s 5$^{\mathrm{th}}$ closest approach, known as a perijove pass (PJ5). PJ5 took place on March 27, 2017 at 8:53 UTC, during which the spacecraft crossed Jupiter's equator at 187$^{\circ}$ W longitude (System III). The data cube presented here was taken under the best atmospheric conditions of our PJ5 observing run (with an average seeing FWHM of 0.786$"$) and was acquired on March 27 between 10:25 and 10:43 UTC. While this particular image cube did not cover the longitude crossed by \textit{Juno}, the atmospheric conditions under which the images were taken allow for absolute photometric calibration of the spectra and for the analysis of specific atmospheric features, such as the SEB outbreak. 

The exposure times for this image cube were chosen such that they were long enough to achieve a high signal-to-noise ratio at a given wavelength but not so long that we would exceed the non-linear count limit for our CCD chip. The change in exposure time over our wavelength range generally reflected the changing response curve of the CCD chip, which was lowest at the bluest and reddest parts of the wavelength range. Our mean exposure time was 1.21 seconds, after using 3.0 seconds from 470-498 nm, 1.7 seconds from 500-558 nm, 1.3 seconds from 560-618 nm, 1.0 second from 620-728 nm, 0.6 seconds from 730-868 nm, and then back up to 1.5 seconds from 870 nm to the end of the cube at 950 nm.

We followed typical telescopic image reduction procedures to convert these images from digital numbers (DN) to physical brightness units. The nature of AOTF data also requires a subtraction of scattered broadband light, which has the advantage of simultaneously subtracting the dark current and CCD bias from the images. These scattered light frames were acquired by imaging the desired targets with no radio-frequency signal applied to the AOTF. 

Before we were able to execute routine flat division procedures or photometrically calibrate the data, we had to correct an additional issue resulting from the CCD used to collect these data: that of optical etaloning or ``fringing" at wavelengths longer than $\sim$720 nm. The CCD we used was back-illuminated, meaning light first travels through the thinned silicon layer of the chip as opposed to through the photodiode array as in front-illuminated CCDs. While such back-illuminated CCDs provide a higher degree of quantum efficiency, silicon grows more transparent at longer wavelengths and can therefore generate internal reflections. When observing with filters that have narrow bandpasses (as with the filters produced by our AOTF), such internal reflections can cause interference within the chip, which results in banded, striped, or mottled fringes across images that must be corrected. The degree of contrast for these fringes is determined by several factors including the filter bandwidth, the spectral energy distribution of the signal within the bandpass, and the signal-to-noise ratio. The difficulty in removing the fringes is exacerbated since Jupiter and the quartz lamps used to take dome flats present differing spectral energy distributions and signal-to-noise ratios, which results in science and calibration images that have different levels of fringe contrast for the same wavelength. This can make removing the fringes with conventional flat division nonviable or can even amplify the fringe contrast at certain wavelengths. 

\begin{figure*}
\centering
\includegraphics[scale=0.7]{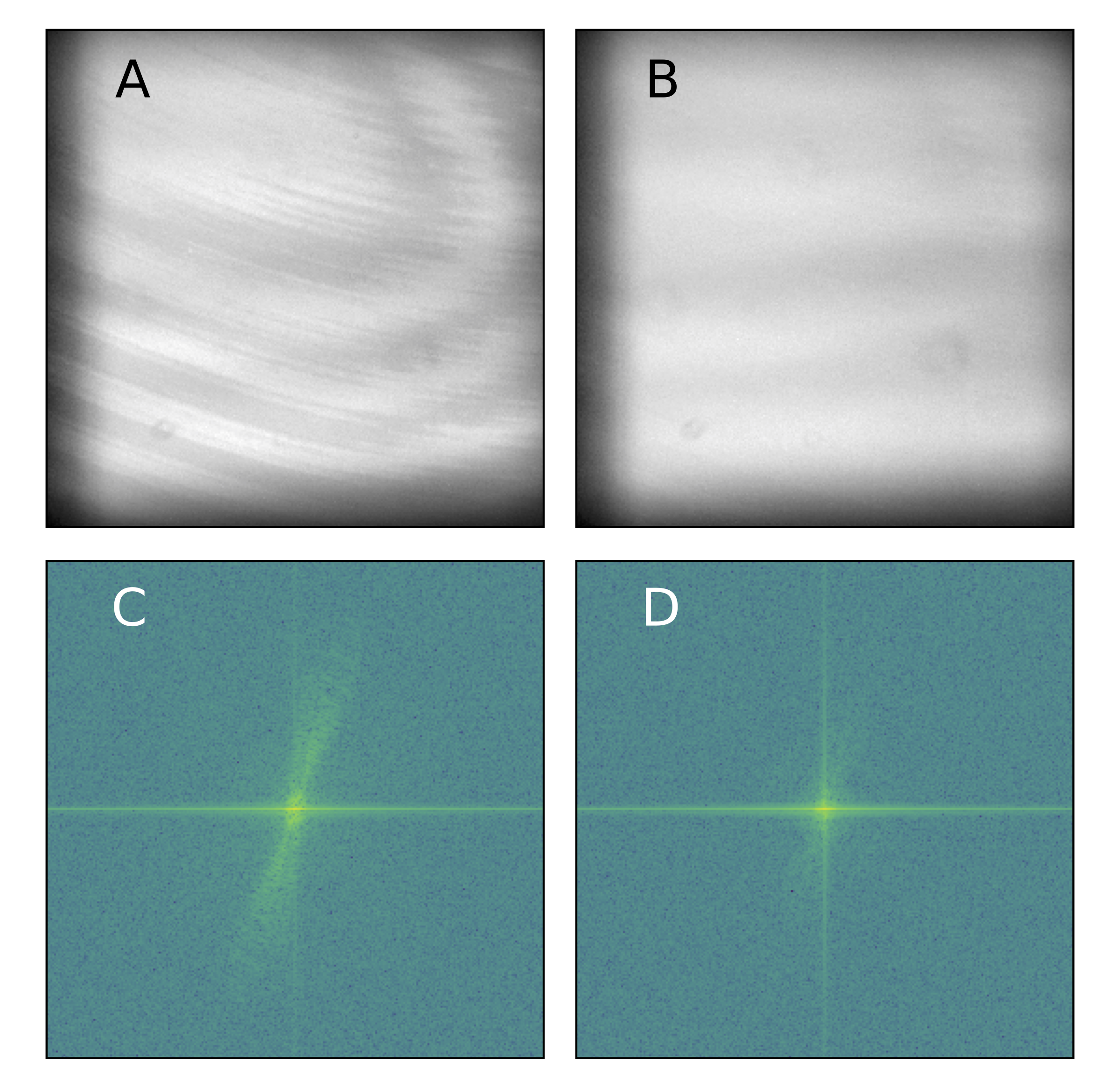}
\caption{Examples of a flat field frame at 920 nm and the same flat in 2D Fourier space, before and after the fringing correction has been applied. Panel A: Uncorrected flat; Panel B: Flat after fringing correction has been applied (horizontal streaks are AOTF transmission variations); Panel C: 2-dimensional Fourier transform of the uncorrected flat in panel A. The fringing signature is present in the power spectrum as a smear from the top right to bottom left quadrant; Panel D: 2-dimensional Fourier transform of the corrected flat in panel B. Here, the signature of the fringing has largely been erased. The residual ``smear" in Panel D is likely the low-contrast horizontal bands remaining in Panel B, which are signatures of the AOTF that we want to retain in the flat.}
\label{fig:fringing}
\end{figure*}

To mitigate this fringing effect, we used both science and calibration images that contained the fringing pattern to iteratively determine the thickness of the CCD chip as a function of pixel position. Once the thickness function of the chip was derived, we  developed ``fringing frames," or images that contained only the signature of the fringing aberration. These fringing frames were normalized such that the pixel values were centered around 1. They were then used, before flat division and the rest of the data reduction pipeline were completed, to divide out fringing patterns much like using a flat-field image to divide out typical pixel-to-pixel variation. In principle, this approach can be used for any CCD chip that produces this aberration. Diagnostic two-dimensional Fourier transforms of fringing and fringing-corrected flats are in Figure \ref{fig:fringing} to demonstrate the efficacy of our technique. For a more thorough discussion of this fringe-removal technique, its application to these data, and the thickness function of the chip used to image the data presented in this work, see \citet{wijerathna_2020}. 

\begin{figure*}
\centering
\includegraphics[scale=0.6]{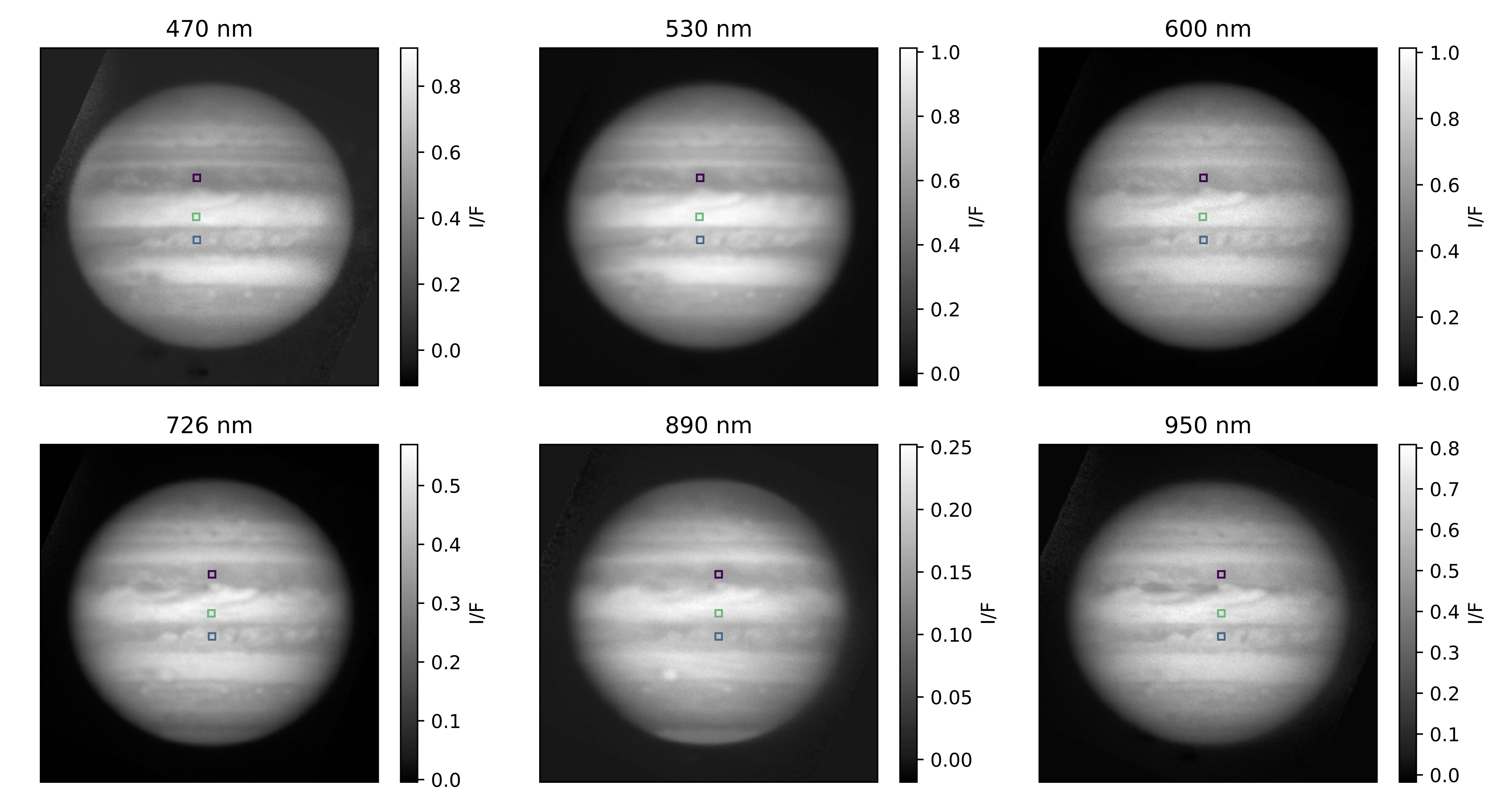}
\includegraphics[scale=0.7]{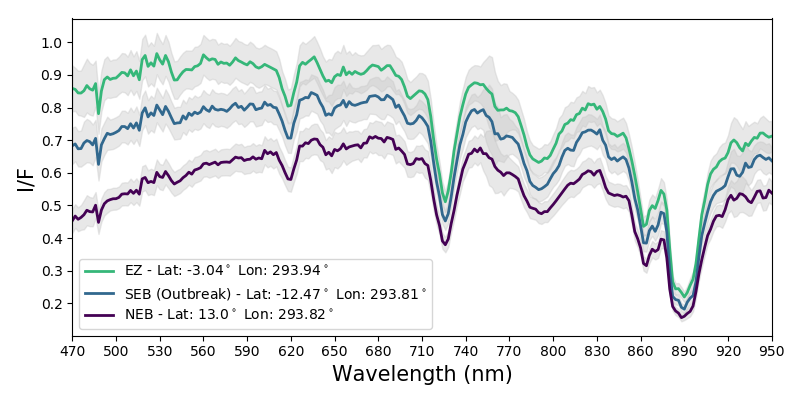}
\caption{Top: Six selected wavelengths/images from the  hyperspectral image cube taken on March 27, 2017, during \textit{Juno}'s 5$^{\mathrm{th}}$ perijove pass. White regions indicate areas of higher reflectivity at a given wavelength. A lack of apparent contrast in the 890-nm image is due to spectral leakage, an issue that arises from the nature of the filter function side lobes. This issue is discussed further at the end of Section 2.2 but is accounted for in our radiative transfer model. The locations from which we extracted spectra outlined in colored boxes in each image. Each colored box corresponds to the same latitude and longitude point, so they move slightly from left to right as the planet rotates $\sim$11 degrees over the course of the image cube. Colors of the 10-by-10 pixel boxes from which we extracted spectra correspond the I/F spectra below. Bottom: I/F spectra of the three cloud bands, with observational uncertainty in the I/F values indicated by the gray shaded region.}
\label{fig:albedospec}
\end{figure*}

In order to geometrically calibrate the image cube, we used the viewing geometry of Jupiter from APO to generate images of a fiducial disk, which we then fit to each Jupiter image in the image cube. These fits provided us with the pixel location of the center of the planet within the frame. The orientation of Jupiter's north pole with respect to the image frame was calculated from telescope pointing and Jupiter's ephemerides. These two pieces of information were used to find the pixel positions of certain latitudes and longitudes of a given point on the planet. Six selected wavelengths and images of Jupiter from the final geometrically calibrated and reduced image cube are shown in Figure \ref{fig:albedospec}.

Once fringe corrections, flat divisions, and geometric calibration were complete, we photometrically calibrated our data. During our PJ5 observations, we imaged q Virgo, a standard A0V star close to Jupiter, over the same range of air masses through which Jupiter passed. Due to their spectral flatness, standard A0V stars can be easily scaled to match high-resolution model Vega spectra when Vega is not available for imaging allowing us to flux-calibrate our data. We conducted aperture photometry of the standard star for each image cube we took, which provided a measure of the star's observed flux in counts per second, $F_{obs}(\lambda)$. We then used this observed flux as a function of the star's air mass, \textit{X}, to fit the following equation:

\begin{equation}
F_{obs}(\lambda) = F_{top}(\lambda) exp(-\tau(\lambda) X)
\label{equation:eq}
\end{equation}

This allowed us to derive both the star's flux at the top of the atmosphere ($F_{top}(\lambda)$) and the optical depth of the atmosphere as a function of wavelength ($\tau(\lambda)$) for the night that we observed. In order to convert counts per second into physical flux units, we calculated a photometric conversion factor, $\beta(\lambda)$, for each observation of q Virgo:

\begin{equation}
    \beta(\lambda) = \frac{F_{V,model}(\lambda)}{F_{top}(\lambda)} = \left[\frac{erg/sec/cm^2/\AA}{DN/sec}\right]
    \label{equation:eq2}
\end{equation}

In Equation \ref{equation:eq2}, the $F_{top}(\lambda)$ associated with q Virgo has been scaled to Vega's magnitude using their respective magnitudes and fluxes. $F_{V, model}(\lambda)$ is a standard Vega flux from high-resolution stellar atmosphere model spectra\footnote{http://kurucz.harvard.edu/stars/vega/}. Once we calculated a photometric conversion factor for each standard star cube, we used the median $\beta(\lambda)$ to calibrate our Jupiter image cube. 

Our final data products, described in the following section, are presented in units of I/F, which is a unitless normalization of albedo that describes the absolute reflectivity of an object. We calculated our Jupiter spectra in units of I/F using the following equation:

\begin{equation}
    I/F(\lambda) = \frac{\pi \beta(\lambda) F_{obs}(\lambda) exp(+\tau(\lambda)X)}{\Omega_{pixel} \frac{F_{S, model}(\lambda)}{r^2_J}}
\end{equation}

where $F_{obs}(\lambda)$ is the spectrum of Jupiter in counts per second as extracted from the image cube, \textit{X} is the air mass of Jupiter at the time an image was taken, $\Omega_{pixel}$ is the solid angle of a NAIC pixel at APO (2.47$\times$10$^{-13}$ sterad), $F_{S, model}(\lambda)$ is a model solar spectrum at 1 AU\footnote{http://kurucz.harvard.edu/stars/sun/}, and $r_J$ is the Jupiter-Sun distance in AU at the time of observation (5.42 AU).

\begin{figure*}
\centering
\includegraphics[scale=0.55]{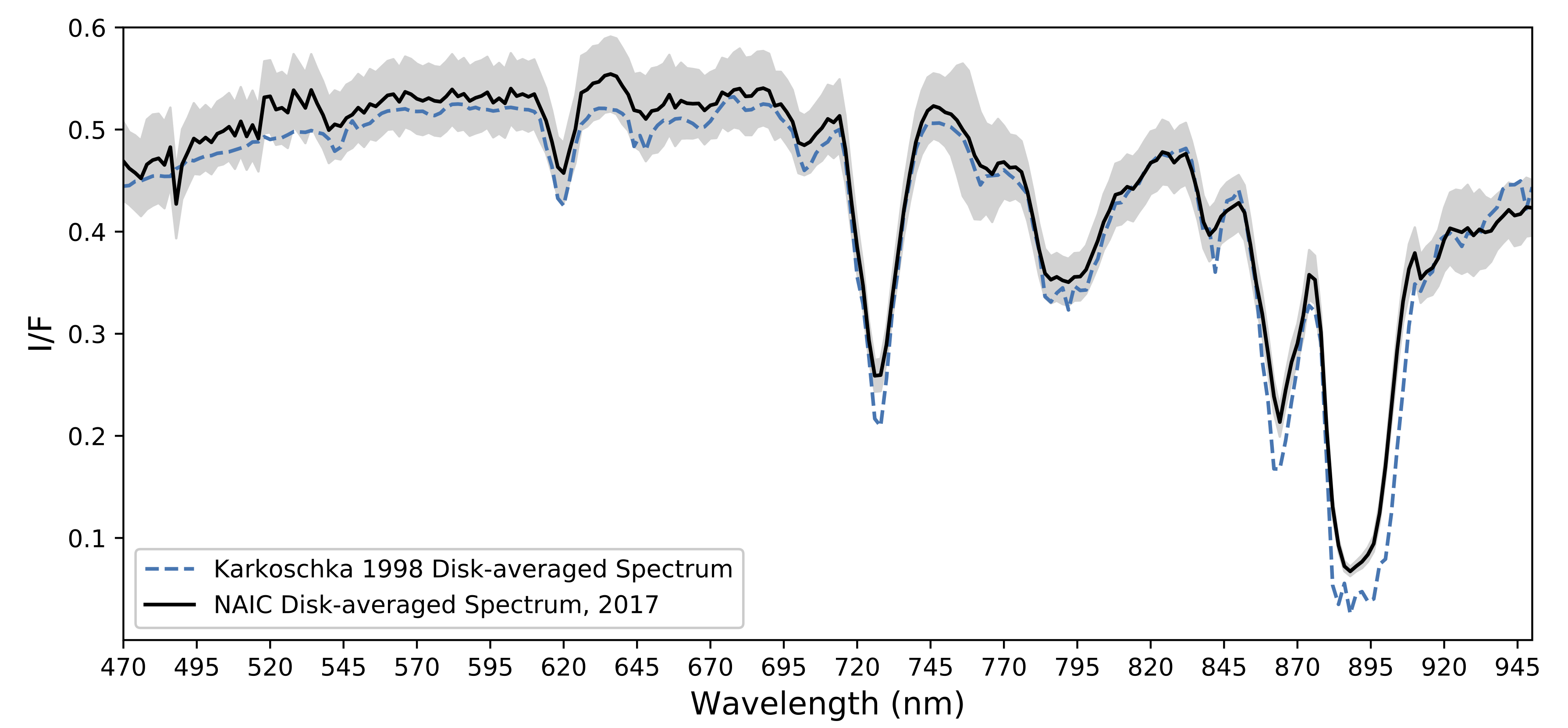}
\caption{Comparison of disk-averaged NAIC spectrum (black with observational error in gray envelope, after a spectral leakage correction has been applied) to that of \citet{kark_1998} (blue dashed line). Discrepancies that remain in absorption bands are likely the result of the combined effects of minor imperfections in our spectral leakage correction and our relatively lower spectral resolution and are not necessarily indicative of any physical changes  in Jupiter’s atmosphere, nor of any fundamental instrumental flaw.}
\label{fig:diskav}
\end{figure*}

As a check on our photometric calibration, we compared our disk-averaged spectrum to that of \citet{kark_1998}, shown in Figure \ref{fig:diskav}. To make these two datasets directly comparable, we applied a correction to the NAIC data in order to account for continuum light leaking into absorption bands via the filters' side lobes (an effect known as ``spectral leakage"). Discrepancies that remain in absorption bands are likely the result of combined effects from minor imperfections in our spectral leakage correction and our relatively lower spectral resolution and are not necessarily indicative of any physical changes in Jupiter's atmosphere, nor of any fundamental instrumental flaw. The shape of NAIC’s filter functions and the resulting spectral leakage effect changes the appearance of our spectra and can potentially decrease our sensitivity to cloud altitudes, particularly in the 890-nm region. The radiative transfer code we use accounts for the shape of our filters and therefore our model atmosphere and retrieved parameters are not affected. We simply point out that these differences in altitude sensitivity between NAIC and the instruments in the other studies cited throughout this work should be kept in mind when comparing our results to other works.

\subsection{Data Products}

We extracted locally-averaged spectra from three of Jupiter's major banded cloud structures: the Equatorial Zone (EZ), North Equatorial Belt (NEB), and an outbreak cloud in the South Equatorial Belt (SEB). For each cloud feature, we took the average of a 10$\times$10 pixel box (without taking seeing into account, approximately 3,336$\times$3,336 km) centered on a chosen latitude and longitude. The EZ spectrum was centered on a planetocentric latitude and System III longitude of -3.04$^\circ$ and 293.94$^\circ$ W; the NEB on -12.47$^\circ$ and 293.94$^\circ$ W; and the SEB on 13.0$^\circ$ and 293.82$^\circ$ W. We intentionally extracted a spectrum from the region of the SEB experiencing an outbreak (storm features thought to be large, highly convective clouds of bright, fresh material whose convective momentum allow them to break through the top of the reddish belt from much deeper pressures) that began erupting from Jupiter's SEB in late 2016 \citep{wong_2018,depater_2019}, in the hope of finding some interesting results from this unique cloud feature. The size of these boxes was chosen to balance an improvement in the signal-to-noise ratio with remaining squarely within the boundaries of the belts and zones, and to also approximately cover two seeing elements (0.786$"\times$2=1.572$"$) along the diagonal of the box. Figure \ref{fig:albedospec} shows the locations of the spectral extraction boxes in 6 of the 241 images in our image cube as they rotate slowly around the planet over the time it took to image the complete cube.

To extract each spectrum, we also had to account for the rotation of the planet over the duration of time it took to obtain an image cube ($\sim$20 minutes, during which Jupiter rotated $\sim$11 degrees). Therefore, we chose locations for each spectrum that evenly straddled the sub-observer point over the course of the image cube (i.e., as the planet rotated). As the cosine of the emission zenith angle (the angle between the observer's line-of-sight and the normal to the spectral footprint) does not deviate far from 1 in these locations, we were able to safely use average viewing geometry values as input to our radiative transfer model. To confirm that averaging viewing geometry quantities over this range of viewing angles would not affect our results, we computed two radiative transfer models: one that used the averaged viewing geometries with a single spectrum, and one that split the spectrum into 10 segments and took into account the slightly different viewing geometries of those segments. We found that the averaged spectrum, like those that we used for this analysis, produced almost the exact same result as the spectrum where the changing viewing geometry was taken into account. The maximum difference in radiance between these two output spectra was 0.65\%, with a median difference of 0.04\%. 

\citet{braude_2020} found that the blue-absorption gradient of the optical spectrum of the NEB was underestimated by the CB model at high zenith emission angles, thereby motivating their derivation of both a new chromophore and a different cloud profile. We conducted the same test by extracting spectra of the NEB at points that were offset by $\pm$60$^\circ$ from the sub-observer longitude and running a retrieval of the CB model parameters. These extracted spectra were different from those we used in the main analysis, as we had to hold the spectral footprints at a single viewing geometry as opposed to a single latitude/longitude point in order to avoid hitting the limb of the planet as Jupiter rotated over the course of the image cube. Regardless, we found that the CB model was able to reproduce these longitudinally-dependent NEB spectra very well at both emission zenith angles, and we did not find the same discrepancy with the blue slope of the spectrum as \citet{braude_2020}.

In this work, we chose to only examine spectra that were near the sub-observer longitude at the time of observation. This choice was motivated both by the fact that we did not detect the blue-gradient issue at high emission angles and by the limited scope of our science objectives for this analysis. In this work, we sought only to validate the CB model parameterization of Jupiter’s uppermost cloud deck, and we could accomplish this goal with a limited range of viewing geometries. It should be noted that while spectra extracted from other emission zenith angles would aid in disentangling some degeneracies between cloud parameters, they would still not completely eliminate them. Consequentially, the cloud characteristics we derive for these locations should be treated with some degree of caution, although a similar degree of caution would still be necessary even when further constraints are applied due to the degeneracies inherent to this spectral range. While additional viewing geometry constraints are not necessary for accomplishing our goals in this work, future analyses of NAIC datasets will include an analysis of spectra over a range of emission angles, thereby taking full advantage of all available constraints and improving our sensitivity to small changes in cloud structure over time and as a function of longitude.

As a check to confirm that Jupiter's atmosphere did not change enough to affect our spectra over the 18 minutes it took to acquire the image cube analyzed herein, we used Outer Planet Atmospheres Legacy (OPAL) data from the Hubble Space Telescope's Wide Field Camera 3 \citep{simon_2015_opal} to inspect the degree of change in Jupiter's clouds over time. Using two OPAL Jupiter maps from April 3, 2017 taken approximately 13 hours apart, we  first took the difference between these two rotations at three representative continuum wavelengths to find maps of the difference in I/F, which enabled us to derive maps of the average change in I/F per minute. Next, we extracted boxes of the average change in I/F per minute from these OPAL difference maps from the same locations as our NAIC spectra. Calculating the average change in I/F over 18 minutes for each location showed that even with these liberal estimates (due to the vastly improved spatial resolution afforded by the Hubble Space Telescope), the degree of change of I/F was always two orders of magnitude below our observational uncertainty for all three locations and wavelengths tested. Therefore, we are confident that if there is even a slight change in reflectivity and therefore cloud structure or color over the 18 minutes it took to acquire our NAIC image cube, we are wholly insensitive to it.

\section{Radiative Transfer Modeling}

In order to use the CB model of Jupiter's atmosphere to measure characteristics of the EZ, NEB, and SEB and to test the universality of the \citet{carlson_2016} chromophore, we utilized the Non-Linear Optimal Estimator for Multi-variate Spectral Analysis (NEMESIS) radiative transfer package \citep{irwin_2008}. This software allowed us to parameterize Jupiter's uppermost cloud deck and methodically retrieve a best-fit synthetic spectrum for each cloud location. 

\subsection{NEMESIS}

NEMESIS is a radiative transfer package designed to be generally applicable to all planetary atmospheres. It has been successfully used to model the atmospheres of our solar system's gas giants \citep{sanz_requena_2019, irwin_2019}, terrestrial planets and moons \citep{teanby_2007,nixon_2013,thelen_2019}, and exoplanets \citep{Krissansen_2018, barstow_2016}. NEMESIS contains two components that work together to produce a best-fit model atmosphere: a radiative transfer code and an optimal estimation retrieval algorithm. The radiative transfer code calculates a synthetic spectrum that would be emitted, reflected, and/or scattered by a model atmosphere, while the optimal estimation retrieval algorithm compares the synthetic and measured spectra in order to iteratively adjust variables and systematically minimize any discrepancy between the two spectra.

NEMESIS can be used to calculate the strength of individual emission or absorption lines (which is highly accurate but very computationally expensive for more than a few lines), or it can be used in ``band mode", which utilizes the method of correlated-k \citep{Lacis_1991} to more efficiently model absorption bands. In our models, we used NEMESIS in band mode to execute multiple-scattering calculations, since scattered and reflected light dominate Jupiter's visible spectrum within our wavelength range. NEMESIS also adds a user-defined error to the observational uncertainty in order to account for sources of error arising from various approximations made during the modeling and retrieval process, such as using the method of correlated-k as opposed to a line-by-line calculation or any uncertainties tied to the reference gas absorption data. We defined this forward-modeling error to be $\sim$1\% of our average radiance over all wavelengths, after finding that in conjunction with our observational uncertainty, it produced reduced $\chi^2$ values on the order of 1. 

\subsection{Model atmosphere}

In order to parameterize Jupiter's atmosphere within NEMESIS, we used the temperature-pressure profile from \citet{seiff_1998} as derived from measurements made by the \textit{Galileo} probe, which is the only available \textit{in situ} measurement of Jupiter's temperature profile. Absorption in Jupiter's visible spectrum is almost entirely dominated by the presence of ammonia, methane, and collisionally-induced absorption from hydrogen and helium gas. Because of this, we eliminated all gases from the model atmosphere except those four. We checked two model atmospheres against each other: one that also contained 9 of the more abundant gases, disequilibrium species, and hydrocarbons in Jupiter's atmosphere (PH$_3$, C$_2$H$_2$, C$_2$H$_4$, C$_2$H$_6$, C$_4$H$_2$, GeH$_4$, AsH$_3$, CO, H$_2$O) and one with just these four. This test revealed a difference in the output radiance of an average of 0.009\% and a maximum difference of 0.037\%, both of which lie well below both the uncertainty in our radiance measurements as well as any additional error introduced by the modeling calculations. We set the deep hydrogen, helium, and methane volume mixing ratios (VMRs) to 0.86, 0.13, and $1.8\times10^{-3}$, respectively, as derived from the \textit{Galileo} entry probe measurements \citep{niemann_1998}. We used an ammonia gas profile as measured in the infrared by \citet{fouchet_1999}, where the deep abundance (below the $\sim$0.7-bar level) is set to a VMR of $2\times10^{-4}$. Above the 0.7-bar level, the abundance decreases rapidly due to reaching saturation equilibrium; above 0.1 bars photodissociation also depletes the NH$_3$ abundance. 

To model gas absorption, we used methane absorption coefficients from \citet{karkoschka_tomasko_2010} and ammonia absorption data from \citet{irwin_2019_2}. Both sources of absorption have already been successfully used in \citet{braude_2020} to model Jupiter's visible spectrum and are the best currently available absorption data in our wavelength regime. Hydrogen- and helium-related collisionally-induced absorption bands were accounted for using data from \citet{borysow_1989_a}, \citet{borysow_1989_b}, and \citet{borysow_2000}. We used our instrument's filter functions and all of these absorption data to calculate k-tables. These k-tables allowed NEMESIS to quickly calculate the amount of ammonia and/or methane absorption measured at a given wavelength, pressure, and temperature. Therefore, the spectral leakage issue discussed previously -- wherein our measured absorption bands are shallower because of our filter shape -- is accounted for by our NAIC-specific k-tables. 

\begin{figure*}
\centering
\includegraphics[scale=0.65]{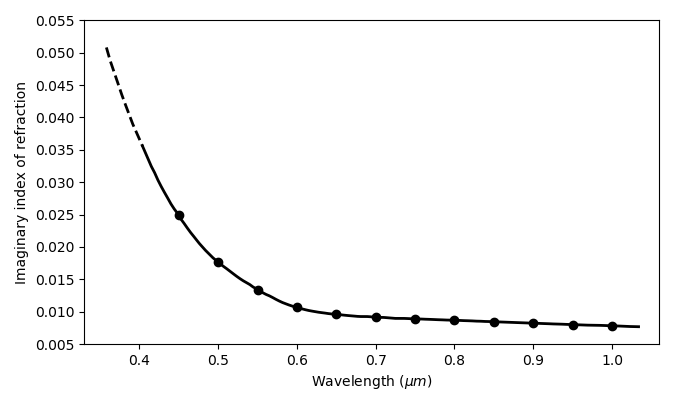}
\caption{Imaginary refractive index spectrum of the ammonia-based chromophore used in our models, as measured by \citet{carlson_2016} (black line). Dots represent the data points used to parameterize the imaginary index within NEMESIS (one every 0.05 $\mu m$, from 0.45-1.0 $\mu m$), and the dashed line shows the usual assumed extrapolation to shorter wavelengths, although this was not measured directly by \citet{carlson_2016}.}
\label{fig:carlson}
\end{figure*}

To parameterize our cloud structure, we used the Cr\`{e}me Br\^ul\'ee (CB) model, which is the most recent and one of the more consistently successful parameterizations of cloud structure for reproducing spectra within our wavelength range, as discussed in Section \ref{sec:intro}. Therefore, we followed the examples of \citet{baines_2019}, \citet{sromovsky_2017}, and \citet{braude_2020} when developing our models and methodology for testing it. The CB model contains a main tropospheric cloud layer, a relatively thin chromophore layer sitting above the tropospheric cloud, and a stratospheric haze layer detached from and above the chromophore layer. Each layer of the model -- the cloud, the chromophore layer, and the stratospheric haze -- has its own optical thickness (denoted $\tau_n$), base ($P_n$) and top pressure ($P_{nT}$), particle radius ($r_n$), and complex index of refraction spectrum. The assumed ammonia abundance profile was parameterized using a simple scaling factor that was the same at all altitudes. While this scaling factor is a simple parameterization of Jupiter's ammonia gas profile, we confirmed that changing the scaling factor by 25\% affected the output spectrum at the same wavelengths and by almost exactly the same amount as changing the deep VMR of the ammonia abundance by 25\%. In other words, increasing or decreasing the deep VMR has the same effect on our modeled spectrum as increasing or decreasing the scaling factor by the same percent amount. This tells us that the upper part of the ammonia profile, where it decreases rapidly with altitude, does not affect the spectrum nearly as much as the deep abundance, otherwise changing our scaling factor would have shown a much larger change in the output spectrum, so we are confident in the ability of this simple parameterization to represent physical changes in the ammonia profile apparent in our data.

For all models, we assumed that the main tropospheric cloud was ammonia-dominated based on evidence of spectrally identifiable ammonia clouds as presented by \citet{baines_2002} and \citet{atreya_2005}. This assumption is also in line with the predictions of the thermochemical equilibrium models in \citet{lewis_1969}, \citet{Weidenschilling_1973}, and \citet{atreya_1999}. We used optical constants for ammonia ice from \citet{martonchik_1984}, leading to the use of a real refractive index of $\sim$1.42 and imaginary index of 0 to model the color of the main cloud. The imaginary index of refraction spectrum of the chromophore is from the ammonia-based coloring agent measured by \citet{carlson_2016}. This laboratory-generated chromophore was made by combining photolyzed ammonia gas with acetylene, resulting in a reddish substance with the imaginary index of refraction spectrum seen in Figure \ref{fig:carlson}. Specifically, we used the \citet{carlson_2016} chromophore spectrum from a sample that was irradiated for 70 hours. It should be noted that while the extrapolated absorption of the \citet{carlson_2016} chromophore extends to 350 nm and our NAIC wavelength range stops at 470 nm, we found that our sensitivity to the location of the shoulder of the chromophore absorption and its slope was sufficient to interpret our results since the slope of the \citet{carlson_2016} chromophore is close to linear shortwards of 500 nm. As mentioned previously, \citet{loeffler_2016} and \citet{loeffler_2018} also presented promising work on a chromophore created by irradiating ammonium hydrosulfide. We did not test this chromophore because of features in this candidate spectrum that are not present in the visible Jovian spectrum, such as the absorption feature at $\sim$600 nm and a lack of strong absorption at wavelengths longer than 500 nm. For the stratospheric haze, we used a complex index of refraction of $1.4+0i$, which is ``a typical value for aliphatic hydrocarbons" \citep{carlson_2016} such as C$_2$H$_2$, C$_2$H$_4$, and C$_2$H$_6$, all of which are abundant in Jupiter's stratosphere \citep{gladstone_1996}. The single-scattering albedo, scattering phase function, and extinction cross-section for all of our cloud layers were calculated given their complex indices of refraction, physical particle sizes, and the use of a Henyey-Greenstein approximation of a Mie-scattering phase function. We assumed a standard gamma size distribution (as defined by \citet{hansen_1974}) of particle size: $n(r)=\mathrm{constant}\times r^{(1-3b)/b}e^{-r/ab}$, where $a$ is the effective radius in microns and $b$ is the dimensionless fixed variance that we held at 0.1. All particle sizes reported in this work are the effective particle size of this distribution. See Table \ref{tab:model_atmosphere} for a summary of the symbols, values, and descriptions of our atmospheric model parameters.

\begin{deluxetable*}{l  l  l c }
\tablecaption{Parameters and their symbols/values as used in this analysis. \label{tab:model_atmosphere}}
\tablewidth{0pt}
\tablehead{
\colhead{Symbol} & \colhead{Parameter description} & \colhead{\textit{A priori} value} & \colhead{Variable? y/n} 
}
\decimalcolnumbers
\startdata
    & \textbf{Main tropospheric cloud} && \\
    $P_1$ & Base pressure & NEB: 3.215 bars; EZ: 2.154 bars; SEB: 4.9 bars & y \\
    $P_{1T}$ & Top pressure & NEB: 0.381 bars; EZ: 0.06 bars; SEB: 0.489 bars & y \\
    $r_1$ & Effective radius of particle & See Tables \ref{tab:sromovsky_particle_sizes} and \ref{tab:braude_particle_sizes} & n \\
    $\tau_1$ & Optical depth & NEB: 16.061; EZ: 13.663; SEB: 25.187 & y \\
    $n_1$ & Complex refractive index & \citet{martonchik_1984} (NH$_3$-dominated)& n \\
    FSH & Fractional scale height & 1.0 & n \\
    & \textbf{Chromophore layer} && \\
    $P_2$ & Base pressure & $P_{1T}$ & y \\
    $P_{2T}$ & Top pressure & 0.9$\times P_2$ & y \\
    $r_2$ & Effective radius of particle & See Tables \ref{tab:sromovsky_particle_sizes} and \ref{tab:braude_particle_sizes} & n\\
    $\tau_2$ & Optical depth & NEB: 0.186; EZ: 0.059; SEB: 0.757 & y \\
    $n_2$ & Complex refractive index & \citet{carlson_2016}; defined from 0.45-10 $\mu m$ every 0.05 $\mu m$ & y and n \\
    & \textbf{Stratospheric haze} && \\
    $P_3$ & Base pressure & 0.01 bar & y \\
    $r_3$ & Effective radius of particle & See Tables \ref{tab:sromovsky_particle_sizes} and \ref{tab:braude_particle_sizes} & n \\
    $\tau_3$ & Optical depth & 0.01 & y \\
    $n_3$ & Complex refractive index & $1.4+0i$ & n \\
    & \textbf{Ammonia abundance profile} && \\
    $f$ & Simple scaling factor & 1.0 & y \\
\enddata

\end{deluxetable*}

\subsection{Sensitivities and degeneracies} \label{sec:sens}

\begin{figure*}
\centering
\includegraphics[scale=0.6]{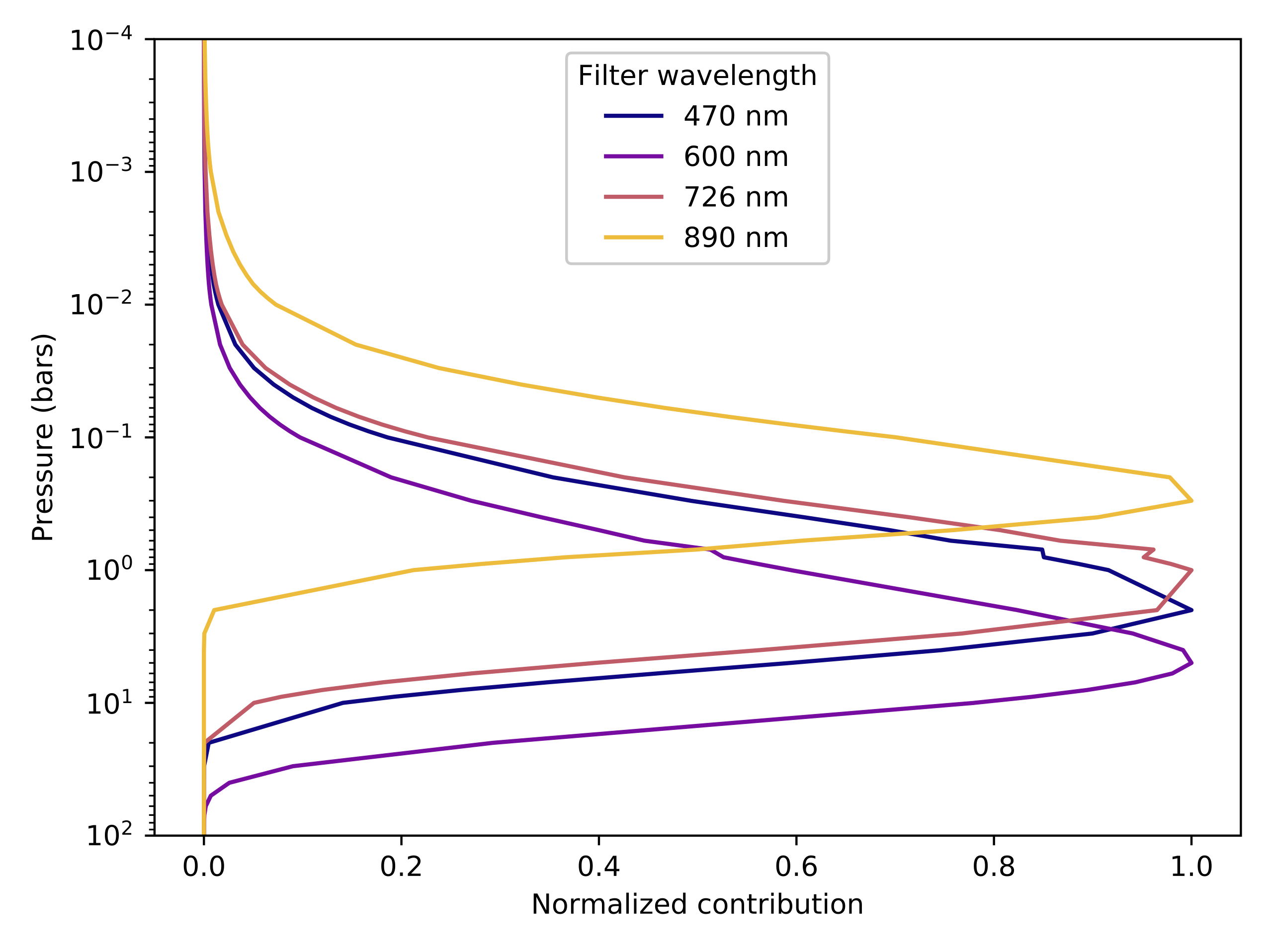}
\caption{Normalized contribution functions for four NAIC filters at representative continuum and methane band wavelengths, showing which altitudes contribute to the observed reflectivity of a given filter in a cloudless Jovian atmosphere.}
\label{fig:cont_functions}
\end{figure*}

In our wavelength range, we are sensitive to Jupiter's cloud structure, including the cloud's base and top pressures, optical depth, particle size, and fractional scale height; the ammonia gas abundance; and the cloud's color by way of the complex index of refraction spectrum. However, these sensitivities overlap at most wavelengths, and we are more sensitive to some parameters than others, making several parameters very nearly degenerate with one another. Because of the degenerate nature of this parameter space, the ranges of sensitivity for a given parameter can change depending on the characteristics of the rest of the atmosphere. We should note that while these parameters are not truly degenerate, which would inhibit our ability to differentiate between them at all, they are very close to being so. For brevity, we will simply refer to various pairs of parameters as degenerate for the remainder of this study. NAIC contribution functions at two continuum wavelengths and two methane band wavelengths are shown in Figure \ref{fig:cont_functions}. These contribution functions illustrate the relative amounts of emergent intensity as a function of pressure for each filter, taking into account Rayleigh scattering and methane gas absorption. Thus, they represent the range of pressures probed by each filter for a cloudless atmosphere. Any aerosols above the contribution function peaks should be readily visible unless obscured by additional overlaying cloud layers.

\begin{figure*}
\centering
\includegraphics[scale=0.48]{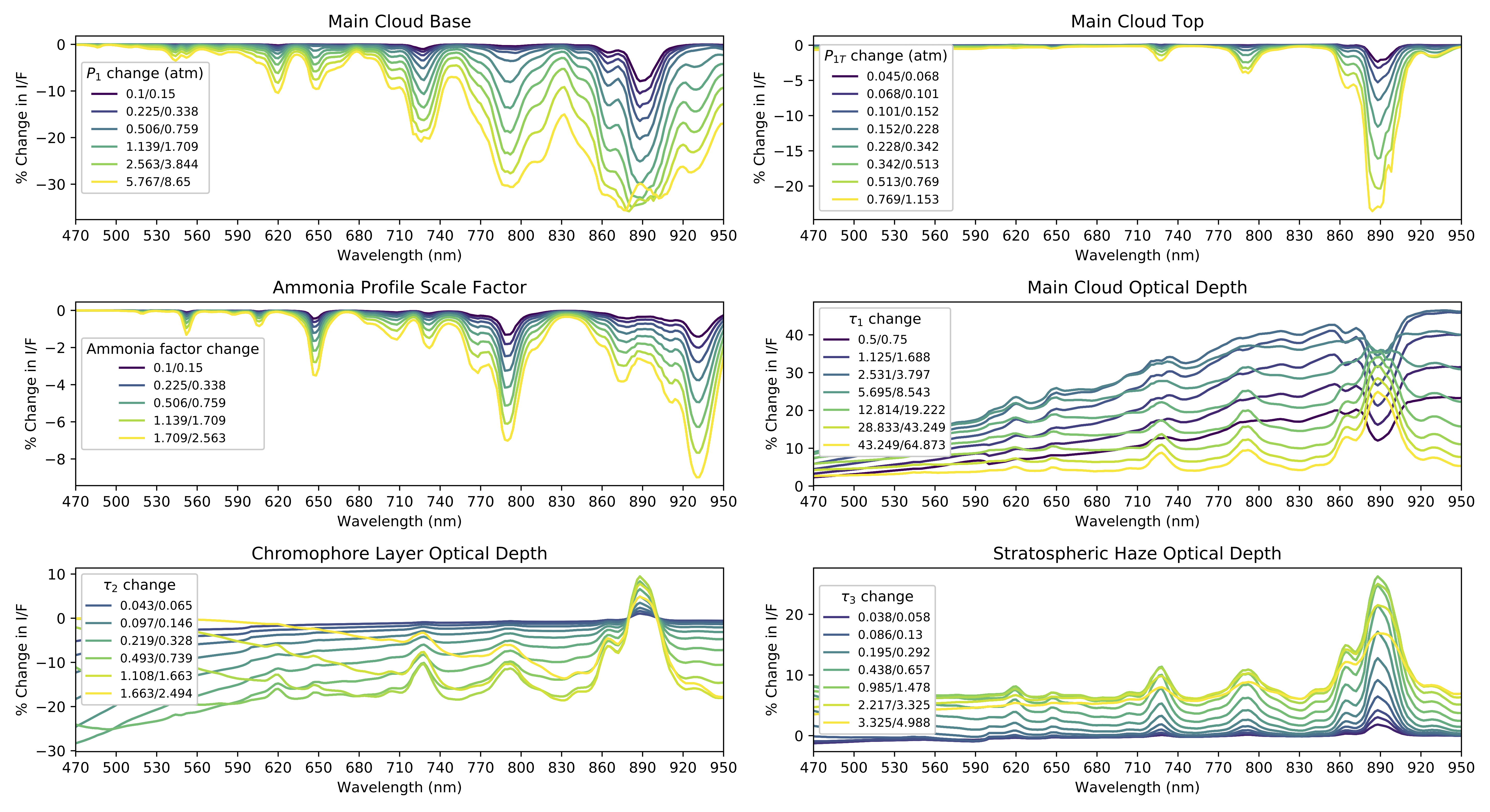}
\includegraphics[scale=0.48]{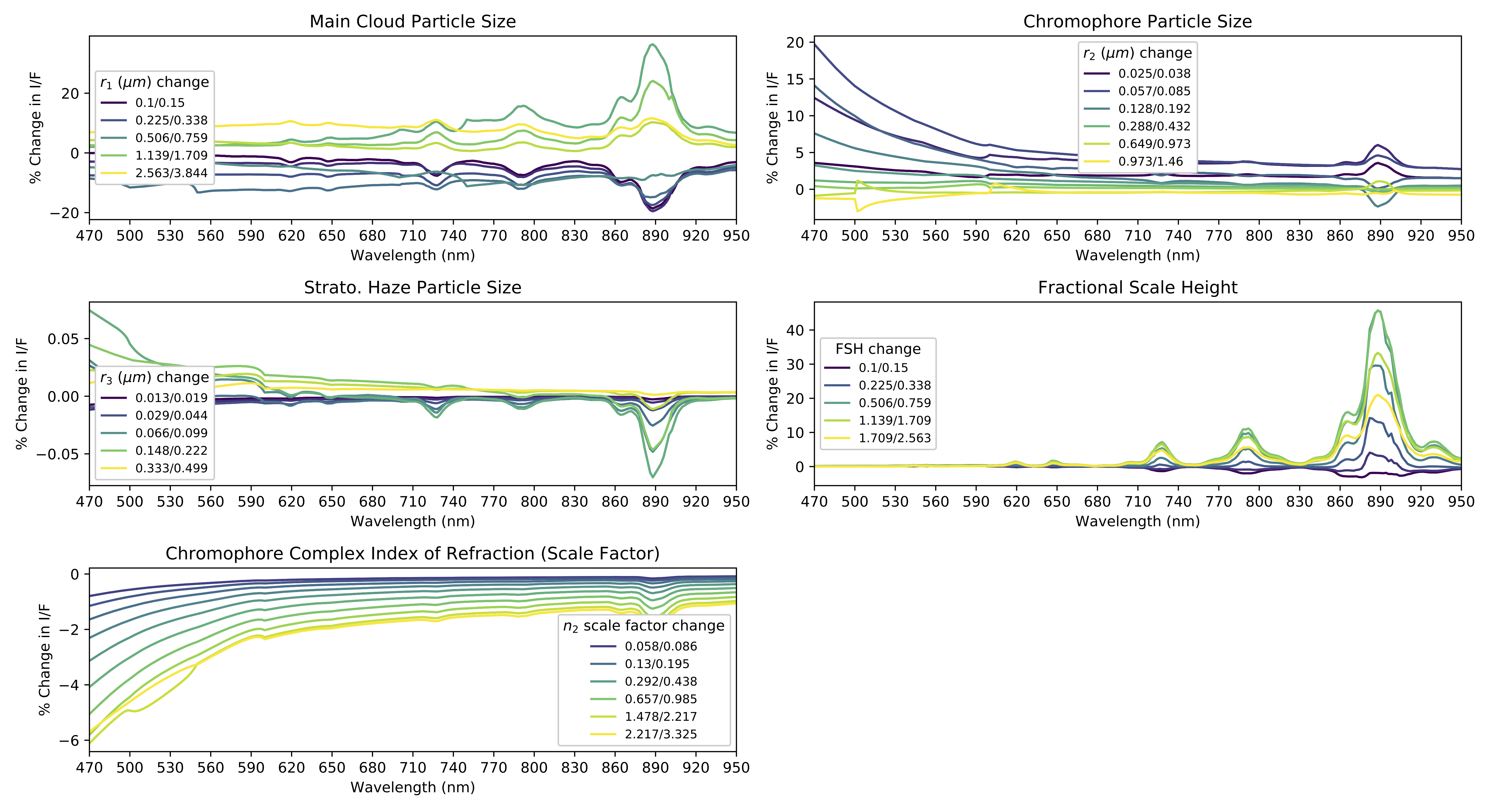}
\caption{Plots of the percent change in I/F for series of 50\% increases in a perturbed variable, for each parameter examined in our model. For example, increasing the chromophore particle size by 50\% from 0.057 microns to 0.085 microns will increase the I/F of the spectrum by $\sim$15\% at 500 nm. To examine our sensitivity to the chromophore imaginary index of refraction spectrum, we multiplied the spectrum from \citet{carlson_2016} by a series of scale factors. For brevity, the legends in each panel contain the first, last, and every other parameter perturbation, but the displayed lines show every incremental change in I/F for consecutive 50\% changes in parameters.}
\label{fig:sensitivities}
\end{figure*}

If sensitivity is defined as the ability of a change in a variable to produce a change in the output spectrum, we sought to better understand our model sensitivities by calculating how the amount I/F (or radiance) would change for incremental increases in each of the parameters examined or varied in this study. To do so, we began with the best-fit atmosphere for the EZ as constrained by multiple viewing geometries from \citet{sromovsky_2017}. We then changed a given parameter by 50\% intervals while leaving all other parameters constant and calculated the output spectrum with a forward model. Comparing adjacent sets of forward models allowed us to find the percent change in the spectra for each 50\% change in parameter. We calculated several percent changes in I/F for each parameter since the percent change in radiance is not always the same for a 50\% change in a given parameter (\textit{e.g.} increasing $r_2$ from 0.038 to 0.057 microns  produces a larger change in radiance than increasing it from 0.28 to 0.43 microns). Figure \ref{fig:sensitivities} illustrates the percent change in I/F (or radiance) for each parameter we analyzed or varied in this study.

Within this degenerate parameter space, identifying a peak, cutoff, or range of sensitivity for a given parameter is nontrivial because those quantities depend on the other atmospheric parameters that are being held constant. For example, we might not be sensitive to a cloud base of 5 bars with a very optically thick cloud, but a cloud with a relatively lower opacity might allow us to detect changes in the location of the cloud base at depth. Regardless, these plots provide us with a general understanding of where our sensitivity peaked, where it began to degrade, and the point at which we should be skeptical of a retrieved result. For example, our peak sensitivity for the main cloud's optical depth lies around 2-8 if we assume the rest of our parameters are constant, but we are still relatively sensitive at both the highest and lowest limits of the optical depths we tested. 

\begin{figure*}
\centering
\includegraphics[scale=0.6]{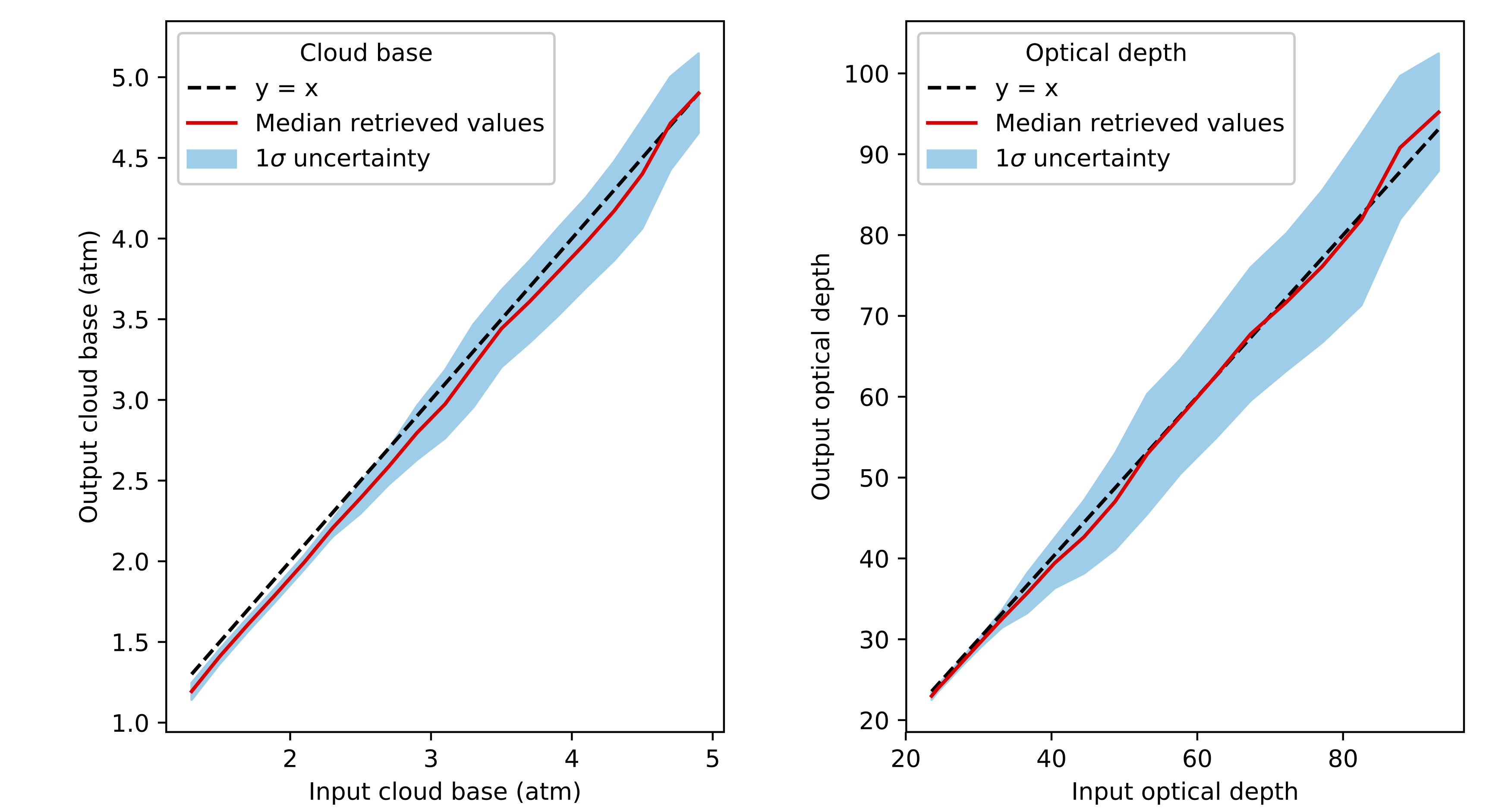}
\caption{These panels show the results of testing the ``retrievability" of degenerate pairs of optical depth and main cloud base. We used NEMESIS to calculate one artificial spectrum for each degenerate pair of optical depth and cloud base, applied 20 sets of random errors to those spectra, and then ran retrievals on all spectra. The panels show how closely correlated the input and retrieved cloud base (left) and input and retrieved optical depth (right) were for all 20 sets of errors. The median of each retrieved pair (red line) and the 1$\sigma$ spread of uncertainty on the results (light blue envelope). If NEMESIS always perfectly retrieved the correct pair of cloud base and optical depth, the median correlation would rest on the y = x line. While NEMESIS does slightly under-estimate the cloud base, the optical depth retrieval is usually accurate.}
\label{fig:degeneracy_test}
\end{figure*}

The degeneracies in Jupiter's visible spectrum can also be read from these percent change plots, such as the optical depth and particle size of the chromophore layer: increasing the particle size and optical depth can result in roughly the same change in radiance as decreasing both of those parameters. Most notably, the main cloud base pressure and optical depth are also positively correlated: a deep, optically thick cloud can produce roughly the same spectrum as a relatively high and optically thin cloud. This degeneracy is one of the most pronounced in this parameterization. In order to test the ability of NEMESIS to retrieve the most degenerate pair of parameters and to ensure we were capable of decoupling them, we computed a series of forward models using pairs of cloud bases and optical depths that produced very similar, albeit distinct, spectra. We next added random noise to these forward models and then used these noisy synthetic spectra as input for a full retrieval of atmospheric parameters. The correlation between input to the forward models, the output from the retrievals, and the median retrieved values and their uncertainty are shown in Figure \ref{fig:degeneracy_test}. We found that while there was some spread in solutions for deeper and thicker clouds, which is to be expected, for the most part NEMESIS successfully re-retrieved the correct combination of optical depth and cloud base. While NEMESIS can differentiate between this pair of parameters, it is important to remember that the degeneracies in this wavelength regime make it nontrivial to perfectly retrieve cloud structures without further constraints from other wavelength regimes or \textit{in situ} measurements.

\begin{figure*}
\centering
\includegraphics[scale=0.6]{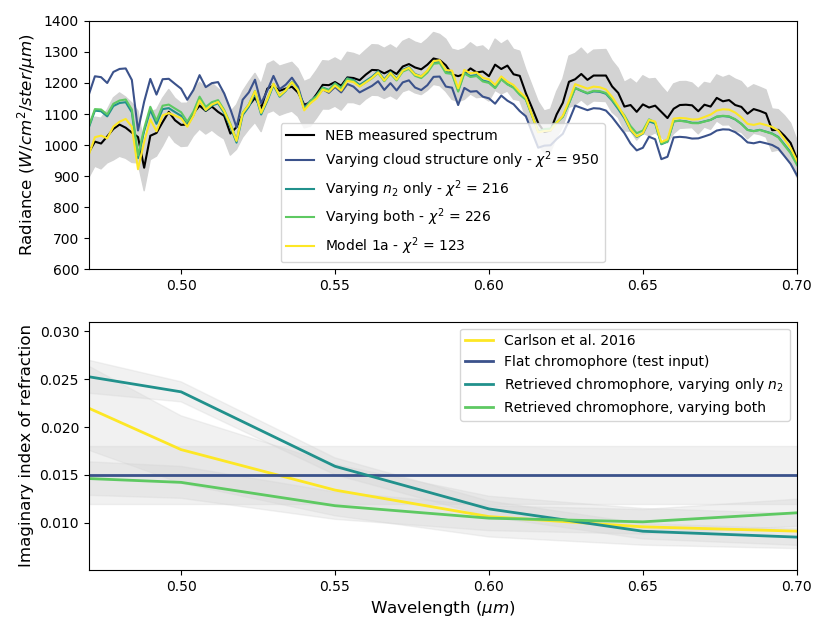}
\caption{The results of using a spectrally flat chromophore as our \textit{a priori} guess to test our sensitivity to the \citet{carlson_2016} chromophore. Top panel: The spectral fits to the spectrum of the NEB when we only allowed cloud structure to vary (indigo line), when we only allowed $n_2$ to vary (teal line), and when we allowed both sets of parameters to vary (green line). Included is the spectral fit result from Model 1a for reference, where we used the \citet{carlson_2016} chromophore as our \textit{a priori} guess and only varied cloud structure. Bottom panel: The \citet{carlson_2016} chromophore for reference (yellow line), the tested \textit{a priori} chromophore imaginary index spectrum (indigo line), the retrieved chromophore spectrum when we allowed only $n_2$ to vary (teal line), and the retrieved chromophore spectrum when we allowed both sets of parameters to vary (green line).}
\label{fig:chromo_flat_test}
\end{figure*}

As an additional test to verify our sensitivity to the color of the chromophore, we used a flat imaginary index spectrum as an \textit{a priori} input for the chromophore instead of the values from \citet{carlson_2016}. The spectral fits and retrieval results from this test are in Figure \ref{fig:chromo_flat_test}. We found that even when we varied different sets of variables that might be able to compensate for its gray color, a spectrally flat chromophore alone could not fit the blue region of the NEB spectrum (0.7 microns and shorter) nearly as well as when we used the \citet{carlson_2016} chromophore as a prior input.

Allowing only cloud structure parameters to vary under the spectrally flat chromophore could not reproduce the data as well as the same model when the \citet{carlson_2016} chromophore was used as input (e.g., Model 1a). When we allowed only the flat $n_2$ to vary, we retrieved a chromophore with the same approximate shape as the one reported by \citet{carlson_2016}. This shows that if only the chromophore is responsible for the broad spectral variances in this region of Jupiter's spectrum, it must have a shape similar to that of the chromophore identified in \citet{carlson_2016}. That is, it must be a broad blue absorber. Additionally, allowing both the spectrally flat chromophore and cloud structure parameters to vary also produced a chromophore with the same general shape, but with less pronounced blue absorption. This difference in the amount of absorption can be accounted for, as some cloud parameters can partly but not completely compensate for brightness variations at those wavelengths. Of these four models of the NEB, the one that used the \citet{carlson_2016} chromophore produced the lowest $\chi^2$ values and the best spectral fit.

Due to the intrinsic degeneracy of this wavelength regime, it cannot be entirely ruled out that some yet-unidentified, exotic chromophore might be able to explain variations in Jupiter's spectrum at these wavelengths. However, this test and previous works have identified the need for a broadly-absorbing blue coloring agent to produce the differences we see across Jupiter's reflectance spectrum \citep{simon_2001a, simon_2001b}. Currently, the most promising candidate for this blue absorber is from \citet{carlson_2016}, so we are confident in our use of this chromophore as an \textit{a priori} guess even when we allow $n_2$ and cloud structure to vary in our models.

\subsection{Methodology}

While NEMESIS is able to differentiate between near-degenerate pairs of optical depth and cloud base, the degeneracies between particle size and other atmospheric parameters are more difficult to separate. NEMESIS is capable of retrieving particle size, although these retrievals can become unstable if the constraints are not carefully tuned. As a result, we needed to assume either a single set of \textit{a priori} particle sizes (although that might bias our results towards a cloud structure reflective of that assumption), or iteratively test different combinations of $r_1$, $r_2$, and $r_3$ in order to determine which combination might provide a best-fit solution. In this work, we followed both approaches in order to avoid biasing our results towards a given particle size distribution as much as possible. 

We used the particle sizes for the EZ, NEB, and SEB as derived by \citet{sromovsky_2017} and we also tested a 3-dimensional discrete grid of particle sizes for each cloud feature, akin to the methodology of \citet{braude_2020}, who also utilized NEMESIS for their analysis. In our first approach, we fixed the particle sizes in the EZ, NEB, and SEB according to  \citet{sromovsky_2017}, which were $r_1$ = 0.586 $\mu m$, $r_2$ = 0.117 $\mu m$, and $r_3$ = 0.1 $\mu m$ for the EZ; $r_1$ = 1.438 $\mu m$, $r_2$ = 0.151 $\mu m$, and $r_3$ = 0.1 $\mu m$ for the NEB; and $r_1$ = 0.836 $\mu m$, $r_2$ = 0.286 $\mu m$, and $r_3$ = 0.1 $\mu m$ for the SEB. In our second approach, we tested a  $6\times6\times3$ grid of particle sizes for each cloud region. We used each possible combination (108 total) from the tested particle sizes of $r_1$, $r_2$, and $r_3$ which were 0.5, 0.75, 1.00, 2.50, 5.00, 7.50 $\mu m$, 0.02, 0.05, 0.1, 0.2, 0.5, 1.0  $\mu m$, and 0.05, 0.1, 0.15 $\mu m$ respectively. In this approach we tested this same grid for each cloud band. These particle sizes are also listed in Tables \ref{tab:sromovsky_particle_sizes} and \ref{tab:braude_particle_sizes}.

Both of these approaches have their own advantages and disadvantages. Using the particle sizes from \citet{sromovsky_2017} tests whether results derived from an analysis of \textit{Cassini} VIMS spectra are consistent with the measurements presented in this study. However, by fixing the particle size to those of a study using observations from $\sim$17 years prior, we could be using particle sizes that no longer represent these cloud regions and which might bias our retrieved cloud structures towards those found in \citet{sromovsky_2017} due to their degeneracy with optical depth. Other earlier works that we discussed in Section \ref{sec:intro} and presented in Table \ref{tab:atm_structure_table} produced a wide variety of particle sizes, so we should not assume that the particle sizes derived by \citet{sromovsky_2017} are the singular true particle sizes and will provide a well-fit spectrum with the cloud structure we see now. Testing a discrete grid of particles allows us to avoid this bias, but we cannot use these grids to produce the real particle size as a retrieval would, but instead can provide a close estimate, albeit one that is more independent of bias we might impose on our results by only assuming sizes from \citet{sromovsky_2017}.

\begin{deluxetable}{cccc}
\tablecaption{Retrieved Particle Sizes from \citet{sromovsky_2017} \label{tab:sromovsky_particle_sizes}}
\tablehead{\colhead{Cloud feature} & \colhead{$r_1$} & \colhead{$r_2$} & \colhead{$r_3$}}
\tablewidth{0pt}
\startdata
    NEB & 1.438 $\mu m$ & 0.151 $\mu m$ & 0.1 $\mu m$ \\
    EZ & 0.586 $\mu m$ & 0.117 $\mu m$ & 0.1 $\mu m$ \\
    SEB & 0.836 $\mu m$ & 0.286 $\mu m$ & 0.1 $\mu m$ \\
\enddata
\tablecomments{Used as inputs for Models 1a and 1b}
\end{deluxetable}

\begin{deluxetable}{ccc}
\tablecaption{Tested Particle Sizes from \citet{braude_2020} \label{tab:braude_particle_sizes}}
\tablehead{ \colhead{$r_1$} & \colhead{$r_2$} & \colhead{$r_3$}}
\tablewidth{0pt}
\startdata
    0.5 $\mu m$ & 0.02 $\mu m$ & 0.05 $\mu m$ \\
    0.75 $\mu m$ & 0.05 $\mu m$ & 0.1 $\mu m$ \\
    1.00 $\mu m$ & 0.10 $\mu m$ & 0.15 $\mu m$ \\
    2.50 $\mu m$ & 0.20 $\mu m$ & \\
    5.00 $\mu m$ & 0.50 $\mu m$ & \\
    7.50 $\mu m$ & 1.00 $\mu m$ & \\
\enddata
\tablecomments{Used as inputs for Models 2a and 2b}
\end{deluxetable}

We also ran two other subsets of models: one where we allowed all cloud structure parameters and the ammonia scale factor to vary, and another where we allowed all cloud structure parameters, the ammonia scale factor, \textit{and} the imaginary index of refraction spectrum for the chromophore layer to vary. If the \citet{carlson_2016} chromophore and CB cloud layering scheme provided accurate fits, then that would be evidence in support of this parameterization. If the \citet{carlson_2016} chromophore, when we held it fixed, was unable to fit our data but provided a more accurate fit when it varied, this would be evidence for a non-universal chromophore, or an entirely different universal chromophore. If both sets of retrievals didn't provide us with accurate fits to the spectrum, that could suggest that the CB model is not a suitable parameterization of Jupiter's uppermost cloud deck. 

We ran an additional set of models to provide some insight to the cloud bands' properties relative to each other by holding cloud bases constant at 3 bars and fixing $r_1$, $r_2$, and $r_3$ to 1.0, 0.15, and 0.1 $\mu m$, respectively. Holding these values constant between our spectra allowed us to compare the variable cloud characteristics and how they differ between cloud features, such as cloud top pressure or the optical depths of any of the layers.

Regardless of our assumptions concerning particle size, we used the cloud structures derived by \citet{sromovsky_2017} as listed in Table \ref{tab:model_atmosphere} as a first assumption unless otherwise noted. We always set the \textit{a priori} error to 25\%, with the exception of the chromophore imaginary index of refraction spectrum, whose \textit{a priori} error was set to 20\% in order to better compare our results to that of \citet{braude_2020}, who allowed the same amount of variation. We found that 25\% was sufficient to allow the fitting algorithm to avoid getting stuck in a local $\chi^2$ minimum but not too high as to allow for ill-fitting or unphysical results. We did not utilize the best-fit cloud structure parameters from \citet{braude_2020} because of their departure from the CB layering scheme. However, we did test some of their best-fit cloud bases in our highly constrained models.

For the sake of simplicity, after this point we will refer to our four sets of models with the following notation, listed here with the main differences between each set:
\begin{itemize}
    \item \textbf{Model 1a}: Did not allow the imaginary index of refraction spectrum of the chromophore ($n_2$) to vary; used the derived particle sizes from \citet{sromovsky_2017}
    \item \textbf{Model 1b}: Allowed $n_2$ to vary and used the derived particle sizes from \citet{sromovsky_2017}
    \item \textbf{Model 2a}: Did not allow $n_2$ to vary and used the particle size grid tested by \citet{braude_2020} plus additional stratospheric haze sizes
    \item \textbf{Model 2b}:  Allowed $n_2$ to vary and used the particle size grid tested by \citet{braude_2020} plus additional stratospheric haze sizes
\end{itemize}

\pagebreak
\section{Results}

\begin{figure*}
\centering
\includegraphics[scale=0.63]{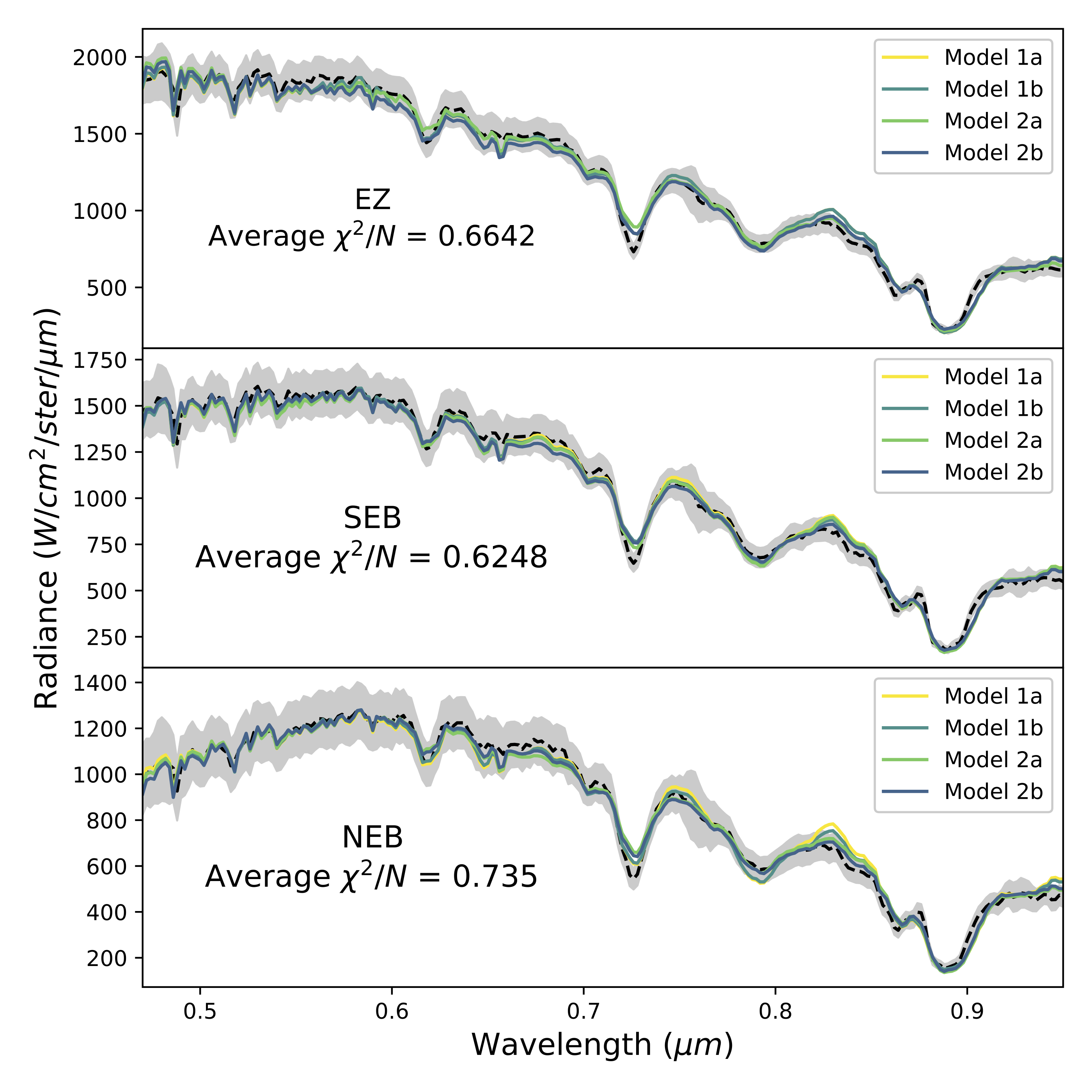}
\caption{The best spectral fits (colored lines) from the four sets of models compared to the data (black dashed line with combined uncertainty from observations and modeling calculations in gray) and the average reduced $\chi^2$ values for each cloud feature.}
\label{fig:all_fits}
\end{figure*}

\begin{figure*}
\centering
\includegraphics[scale=0.55]{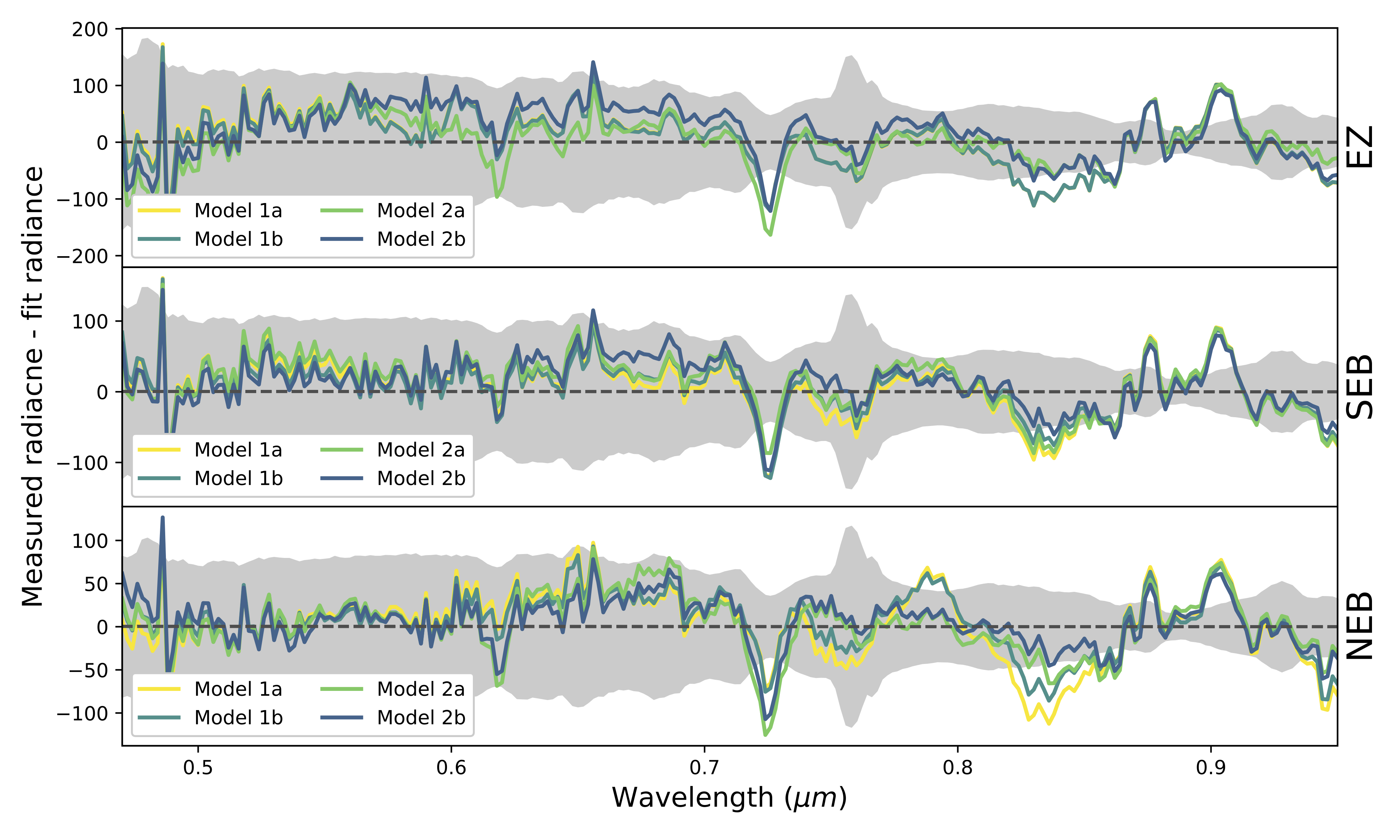}
\caption{The residuals from the best spectral fits from each of our four sets of models, with colors corresponding to those in Figure \ref{fig:all_fits}. Gray envelopes represent the combined observational and forward model uncertainty. The residuals that are consistent between models are discussed in Section \ref{Discrepancies}.}
\label{fig:all_residuals}
\end{figure*}

We found that regardless of the prior assumptions we made, all four sets of models produced very similar fits to the data, all with reduced $\chi^2$ values below but on the order of 1. In Figure \ref{fig:all_fits}, we show the four best-fit retrieved spectra for each cloud band, and Figure \ref{fig:all_residuals} shows the corresponding residuals. For models 2a and 2b, where we tested the particle grid, we present models from the 108 size combinations that produced the best-fit results. While the spectral fits themselves are similar despite the different prior assumptions behind them, those assumptions affected the retrieved cloud structure parameters, which are distinct from each other across the sets of models. Lists of the retrieved cloud structure and ammonia parameters for each best-fit model can be found in Tables \ref{tab:411_solutions}-\ref{tab:422_solutions}. See Figures \ref{fig:41_complexn} and \ref{fig:42_complexn} for plots of the retrieved imaginary index of refraction spectra for our different prior particle size assumptions. Throughout this section, it should be remembered that while we use the reduced $\chi^2$ to quantify the goodness of fit of a given model, these values are more significant for models with fewer free parameters, such as Models 1a and 2a, when we did not allow $n_2$ to vary.

\subsection{Chromophore retrievals}

\begin{figure*}
\centering
\includegraphics[scale=0.6]{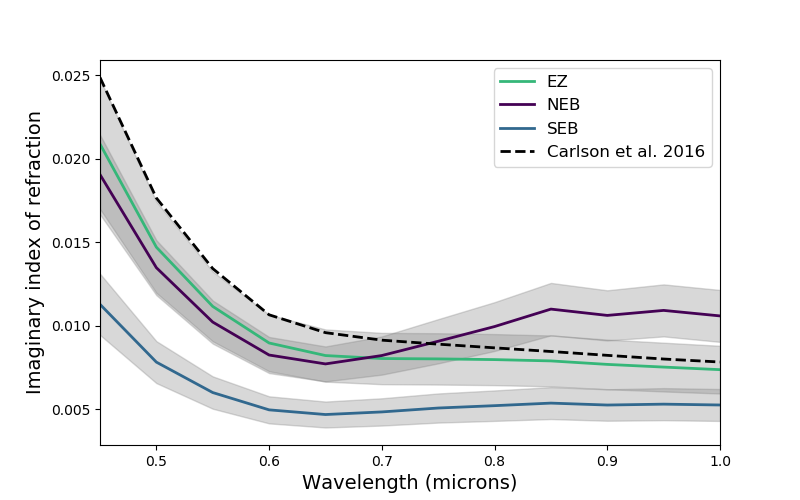}
\caption{Results from the chromophore imaginary index of refraction spectrum retrievals for each cloud band in Model 1b. The gray regions are the uncertainty in the retrieved imaginary index at each wavelength.}
\label{fig:41_complexn}
\end{figure*}

\begin{figure*}
\centering
\includegraphics[scale=0.6]{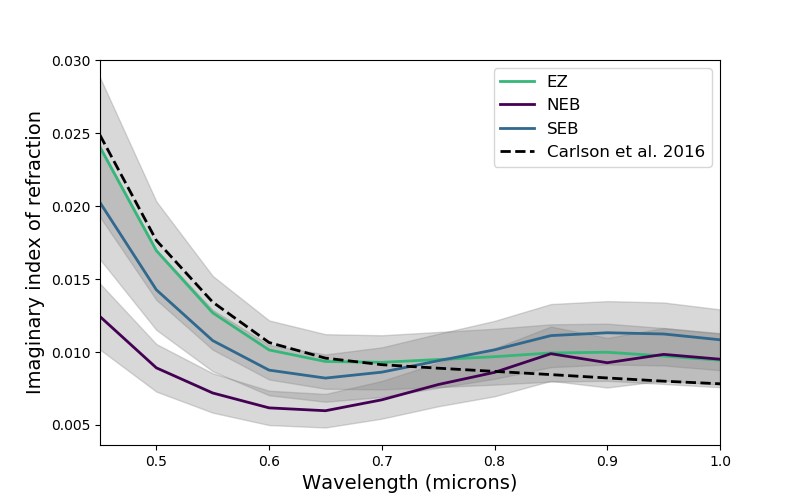}
\caption{Results from the chromophore imaginary index of refraction spectrum retrievals for each cloud band in Model 2b. The gray regions are the uncertainty in the retrieved imaginary index at each wavelength.}
\label{fig:42_complexn}
\end{figure*}

Allowing the imaginary index of refraction of the \citet{carlson_2016} chromophore ($n_2$) to vary did not significantly improve the goodness-of-fit from when we held it fixed. \citet{braude_2020} found that the \citet{carlson_2016} chromophore failed to fit the blue slope ($\lambda < \sim600$ nm) of their NEB spectrum at high emission zenith angles. However, we found that the \citet{carlson_2016} chromophore fit this blue region of both our high- and low-zenith-angle spectra just as well, if not better, than when we allowed it to vary. See Figures \ref{fig:41_complexn} and \ref{fig:42_complexn} for the final $n_2$ spectra for Models 1b and 2b, respectively. 

In Model 1b, we retrieved a much more red-absorbent chromophore over the NEB, while the EZ chromophore layer's spectrum remained close to the original, and a much less absorbent chromophore was retrieved across all wavelengths for the SEB outbreak region. This NEB $n_2$ result is likely the response from NEMESIS  to the residual at $\sim$830 nm in Model 1a, while the $n_2$ result in the SEB is potentially an outcome of the bright outbreak happening in this cloud band at the time of our observations. 

In Model 2b, when we allowed $n_2$ to vary over the particle size grid, we saw more similarities between $n_2$ spectra across the cloud bands than in Model 1b. Each cloud band showed a more absorbent chromophore at redder wavelengths, and a less absorbent one at bluer wavelengths. If Model 2b had produced a significantly better spectral fit than Model 2a, when we did not allow $n_2$ to vary, this would be evidence for a new universal chromophore with less and more absorbency at bluer and redder wavelengths, respectively. Furthermore, if Model 1b had produced a better spectral fit than Model 1a, that would also be evidence that a chromophore other than the \citet{carlson_2016} chromophore might be required. However, the models that held the \citet{carlson_2016} chromophore fixed fit our spectra just as well as when we allowed $n_2$ to vary, despite adding further degrees of freedom to the parameter space.

While it is nontrivial to measure due to the 11 degrees of freedom within the parameterization of our $n_2$ spectra, we can look again at Figure \ref{fig:sensitivities} and see that there seems to be a positive correlation between boosting the absorbency of the $n_2$ spectrum by some universal scaling factor and increasing the particle size of the chromophore. However, since the data points in the $n_2$ spectra were allowed to vary somewhat independently of each other, it is difficult to say whether or not that same degeneracy carried over into our retrieved $n_2$ values.

\subsection{Particle Sizes}

We found that various prior particle size assumptions did not significantly affect the quality of the spectral fit, with the exception of the SEB outbreak region which had reduced $\chi^2$ values above 2.0 in 65 of the 108 particle sizes we tested. Of the $6\times6\times3$ size grid that we tested, the best-fit values between cloud features had no discernible pattern. The only model layer that had consistent best-fit particle sizes between cloud bands was the chromophore layer in Model 2b. In Model 2a, the best-fit main cloud particle size ($r_1$) values were 5.0 $\mu m$, 5.0 $\mu m$, and 0.75 $\mu m$ for the NEB, SEB, and EZ respectively. The chromophore particle size, $r_2$, measured 0.02, 0.2, and 0.5 $\mu m$ for the NEB, EZ, and SEB, and the stratospheric haze size, $r_3$, was different for each cloud band, with 0.1 $\mu m$, 0.15 $\mu m$, and 0.05 $\mu m$ fitting the NEB, EZ, and SEB respectively. In Model 2b, when we allowed $n_2$ to vary, the NEB $r_1$ size was much smaller at 0.5 $\mu m$; the EZ and SEB both measured 2.5 $\mu m$. In this model, $r_2$ was the same size for all cloud bands at 0.02 $\mu m$. $r_3$ was different for each cloud for Model 2b as well, with 0.1 $\mu m$, 0.05 $\mu m$, and 0.15 $\mu m$ fitting the NEB, EZ, and SEB. 

\begin{figure*}
\centering
\includegraphics[scale=0.6]{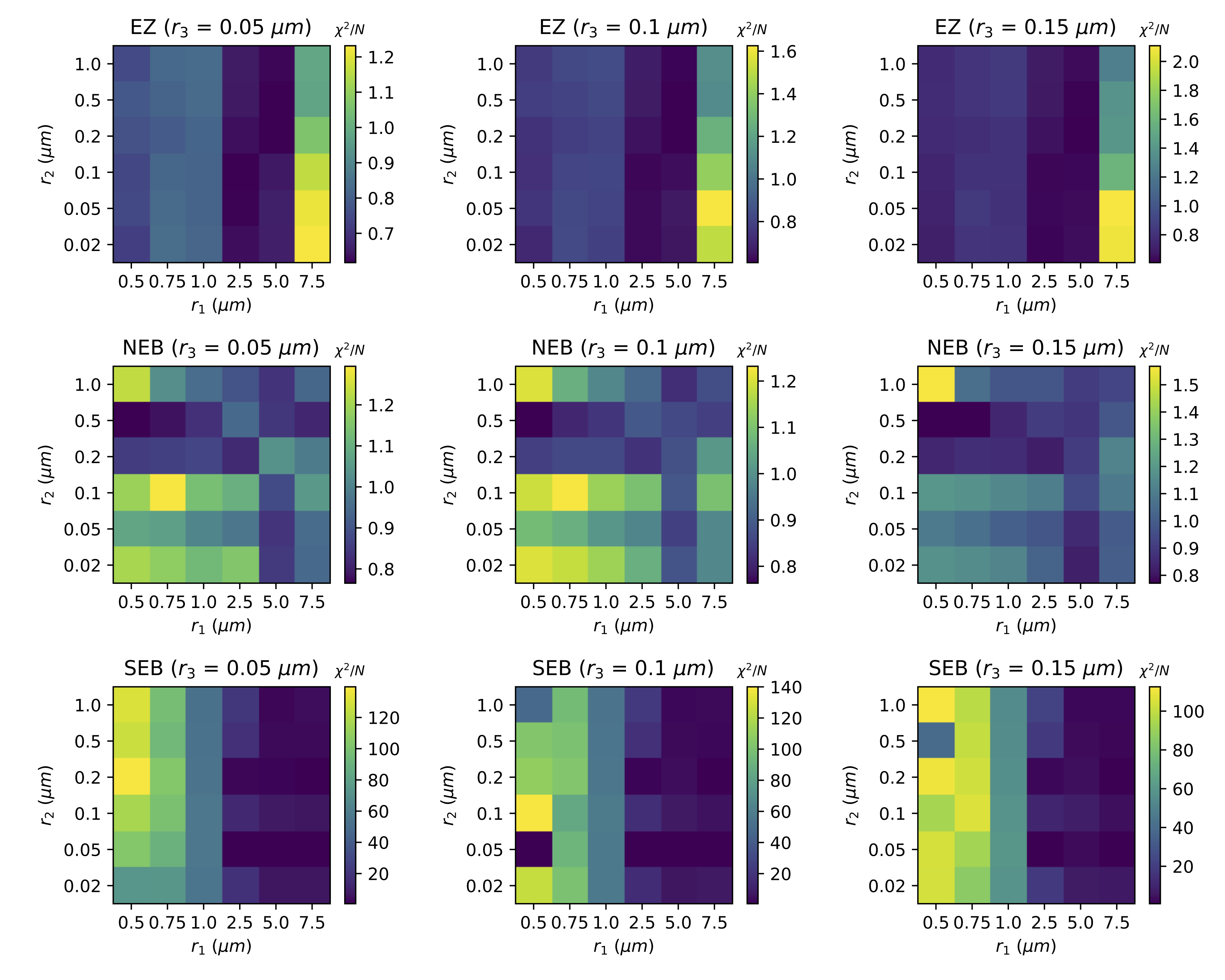}
\caption{Grids of reduced $\chi^2$ results as a function of chromophore and main cloud particle size from Model 2a. There is one panel for each cloud band we modeled, and one for each of the three stratospheric haze particle sizes we tested. These results are from Model 2a, when we did not allow the chromophore imaginary index of refraction to vary. The dashed white lines indicate the particle sizes that we tested, and the darkest parts of the plots indicate regions with relatively low reduced $\chi^2$ values.}
\label{fig:421_grid}
\end{figure*}

\begin{figure*}
\centering
\includegraphics[scale=0.6]{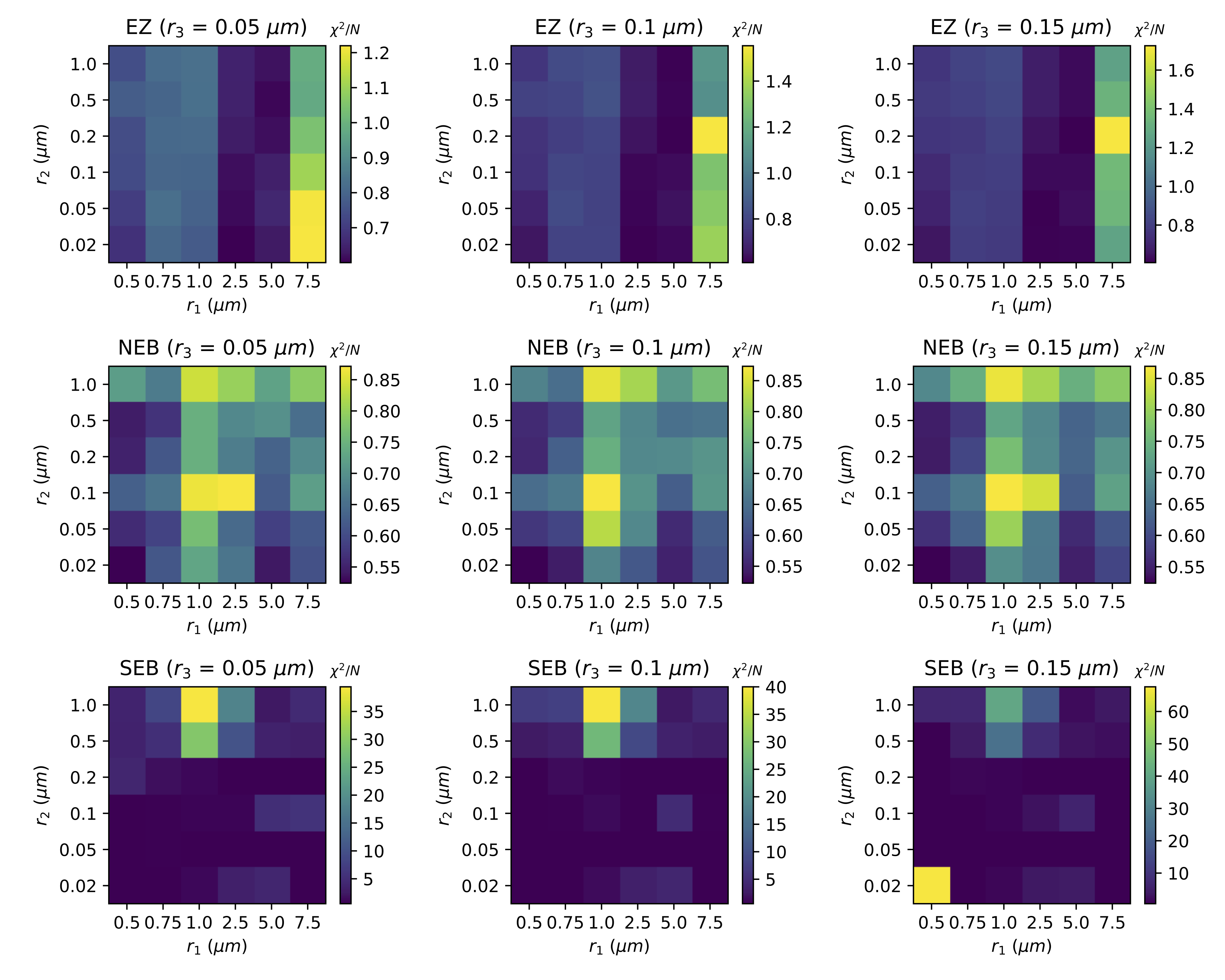}
\caption{Grids of reduced $\chi^2$ results as a function of chromophore and main cloud particle size from Model 2b. There is one panel for each cloud band we modeled, and one for each of the three stratospheric haze particle sizes we tested. These results are from Model 2b, when we did allow the chromophore imaginary index of refraction to vary. The dashed white lines indicate the particle sizes that we tested, and the darkest parts of the plots indicate regions with relatively low reduced $\chi^2$ values.}
\label{fig:422_grid}
\end{figure*}

While there was not much consistency in the best-fit particle sizes among cloud bands for Models 2a and 2b, there were patterns of the reduced $\chi^2$ values within the grids that we tested. For plots of reduced $\chi^2$ as a function of particle size for Models 2a and 2b, see Figures \ref{fig:421_grid} and \ref{fig:422_grid}. The variation between the three different stratospheric haze particle sizes is subtle but sometimes shifted the location of a $\chi^2$ minimum within the main cloud/chromophore layer grid. 

In Figure \ref{fig:421_grid}, showing results from Model 2a, the EZ had a consistent valley of reduced $\chi^2$ values at the tested $r_1$ = 2.5 and $r_1$ = 5 $\mu m$ regardless of the values of $r_2$ or $r_3$, showing that in our models a range of $r_2$ and $r_3$ values could likely fit our EZ cloud as long as $r_1$ was within the 2.5-5 $\mu m$ range. In the NEB, where the chromophore layer is presumably optically thicker, we are likely more sensitive to changes in the chromophore particle size due to its increased abundance. We can indeed see more variation in reduced $\chi^2$ as a function of $r_2$ in these plots. The CB model was consistently able to fit the NEB for almost all particle size combinations that we tried, except for relatively small peaks in reduced $\chi^2$ values around very small $r_2$ and $r_1$ values and at the largest $r_2$ and smallest $r_1$ values. In contrast, the quality of spectral fits for the SEB outbreak region were often much worse, with more than half of the 108 particle size combinations that we tested producing a reduced $\chi^2$ value above 2.0. This shows that in order to use the CB model to accurately model this outbreak cloud, the goodness-of-fit is much more dependent on our prior selection of particle sizes than for other cloud features.

In Figure \ref{fig:422_grid}, we show the same reduced $\chi^2$ results over our particle grids from Model 2b, when we allowed $n_2$ to vary. The same valley of low $\chi^2$ values is present in the EZ around the same range of $r_1$ from 2.5-5 $\mu m$, and the chromophore size is similarly unconstrained for a given $r_1$ value. The low NEB $\chi^2$ values show that again, all tested particle sizes provided an accurate fit to the spectrum. Again, the SEB outbreak region generally has high $\chi^2$ values across the ranges of sizes tested. While the maximum reduced $\chi^2$ was lowered by an order of magnitude, allowing $n_2$ to vary in Model 2b did not improve the goodness-of-fit to the point where our SEB outbreak region can be fit regardless of particle size assumption as in the NEB or EZ.

\subsection{Retrieved cloud structures}

The results for the retrieved cloud structure and ammonia parameters for Models 1a, 1b, 2a, and 2b can be found in Tables \ref{tab:411_solutions}, \ref{tab:412_solutions}, \ref{tab:421_solutions}, and \ref{tab:422_solutions}, respectively. While cloud structure results varied between individual models, there were some similarities in results between Models 1a and 1b and between 2a and 2b. In Models 1a and 1b, when we held particle sizes fixed at those listed in Table \ref{tab:sromovsky_particle_sizes}, we found that cloud base and top pressures for the NEB, EZ, and SEB outbreak region were similar regardless of allowing $n_2$ to vary or not. The main differences between allowing $n_2$ to vary or not can be seen in the changes in main cloud optical depth ($\tau_1$) and chromophore optical depth ($\tau_2$) for the SEB outbreak region. Not allowing $n_2$ to vary produced a very high $\tau_1$ value of $81.888\pm14.887$ for the SEB outbreak, almost twice the value we found for the EZ. Interestingly, Model 1b showed that letting $n_2$ vary almost doubled $\tau_2$ above the SEB outbreak region but lowered its $\tau_1$ to a more reasonable value. The NEB $\tau_1$ value also increased from Model 1a to 1b. It is clear that when the SEB outbreak and the NEB models produced clouds with deeper bases the optical depth was also higher, which is to be expected as a result of the cloud base/optical depth degeneracy, but the $\tau_1$ bar$^{-1}$ value for the SEB in Model 1a (16.73 bar$^{-1}$) is considerably higher than the NEB at a similar altitude in Model 1b (7.57 bar$^{-1}$). This points to the fact that there might be an issue other than parameter degeneracy in the models that is producing this exceptionally high $\tau_1$ value for the SEB outbreak in Model 1a.

Between Models 2a and 2b, we see some similar patterns, such as the extremely high $\tau_1$ value in the SEB outbreak region when we do not allow $n_2$ to vary, but there are also some new issues arising from the degeneracy between particle sizes and optical depths for the different cloud layers. The main cloud bases are almost the same for the NEB and SEB outbreak between Models 2a and 2b, but the base of the EZ moved almost 1 bar deeper when we allowed $n_2$ to vary and when the best-fit particle size was cut in half. Model 2a showed some much optically thinner main clouds for the NEB and EZ than we retrieved in Models 1a and 1b, but that can be traced back to the degeneracy between those sizes and the optical depths of those layers, seeing as the NEB $r_1$ sizes shrink by an order of magnitude between Models 2a and 2b. Similarly, it is difficult to draw conclusions from the relative chromophore optical depths in Model 2a since the best-fit SEB outbreak $r_2$ size is 25$\times$ that of the NEB and its corresponding $\tau_2$ reflects that. It is likely not the case that the SEB outbreak cloud, which we can see is brighter at all wavelengths and lacks the same steepness of the blue spectral slope as the NEB spectrum, has a more optically thick chromophore layer than a redder region of Jupiter's clouds. Model 2b, however, has constant $r_2$ values between the cloud bands and those $\tau_2$ values meet our expectations when compared to each other. That is, the EZ has the most optically thin chromophore layer, followed by the SEB outbreak, followed by the reddish NEB band. In all models, both $P_3$ and $\tau_3$ never varied far from their \textit{a priori values}, confirming our lack of sensitivity to the stratospheric haze when compared to all other parameters. A further analysis of these retrieved parameters and whether or not these results are physical can be found in Section \ref{summary}.

\begin{deluxetable*}{cccc}
\tablecaption{Retrieved atmospheric parameters from Model 1a\label{tab:411_solutions}}
\tablewidth{0pt}
\tablehead{\nocolhead{} & \colhead{NEB} & \colhead{EZ} & \colhead{SEB (Outbreak)}}
\startdata
    $r_1$ &  1.438 $\mu m$ & 0.586 $\mu m$ & 0.836 $\mu m$ \\ 
    $P_1$ &  4.281  $\pm$  0.391  bar  &  3.017  $\pm$  0.253  bar &  4.893  $\pm$  0.645  bar \\ 
    $ \tau_1$ &  26.754  $\pm$  3.374  &  45.948  $\pm$  5.231  &  81.888  $\pm$  14.887  \\ 
    $P_{1T}$ &  0.119  $\pm$  0.023  bar  &  0.043  $\pm$  0.01  bar  &  0.122  $\pm$  0.025  bar  \\ 
    $r_2$ &  0.151 $\mu m$ &  0.117 $\mu m$ & 0.286 $\mu m$ \\
    $ \tau_2$ &  0.272  $\pm$  0.011  &  0.037  $\pm$  0.006  &  0.421  $\pm$  0.024  \\ 
    $r_3$ &  0.1 $\mu m$ &  0.1 $\mu m$ & 0.1 $\mu m$ \\
    $P_3$ &  0.010  $\pm$  0.003  bar  &  0.010  $\pm$  0.003  bar  &  0.010  $\pm$  0.003  bar  \\ 
    $ \tau_3$ &  0.010  $\pm$  0.003  &  0.009  $\pm$  0.002  &  0.009  $\pm$  0.002  \\ 
    $f$ &  1.060  $\pm$  0.140  &  1.389  $\pm$  0.209  &  1.331  $\pm$  0.199  \\ 
    $\chi^2/N$ & 0.954 & 0.730 & 0.703 \\
\enddata
\end{deluxetable*}

\begin{deluxetable*}{cccc}
\tablecaption{Retrieved atmospheric parameters from Model 1b\label{tab:412_solutions}}
\tablewidth{0pt}
\tablehead{\nocolhead{} & \colhead{NEB} & \colhead{EZ} & \colhead{SEB (Outbreak)}}
\startdata
    $r_1$ &  1.438 $\mu m$ & 0.586 $\mu m$ & 0.836 $\mu m$ \\ 
    $P_1$ &  4.651  $\pm$  0.566  bar  &  3.037  $\pm$  0.259  bar &  4.292  $\pm$  0.576  bar \\ 
    $ \tau_1$ &  35.247  $\pm$  6.161  &  46.552  $\pm$  5.336  &  66.978  $\pm$  12.629  \\ 
    $P_{1T}$ &  0.142  $\pm$  0.026  bar  &  0.043  $\pm$  0.01  bar  &  0.154  $\pm$  0.029  bar  \\ 
    $r_2$ &  0.151 $\mu m$ &  0.117 $\mu m$ & 0.286 $\mu m$ \\
    $ \tau_2$ &  0.367  $\pm$  0.031  &  0.041  $\pm$  0.007  &  0.727  $\pm$  0.080  \\ 
    $r_3$ &  0.1 $\mu m$ &  0.1 $\mu m$ & 0.1 $\mu m$ \\
    $P_3$ &  0.010  $\pm$  0.003  bar  &  0.010  $\pm$  0.003  bar  &  0.010  $\pm$  0.003  bar  \\ 
    $ \tau_3$ &  0.010  $\pm$  0.003  &  0.010  $\pm$  0.002  &  0.010  $\pm$  0.002  \\ 
    $f$ &  1.039  $\pm$  0.158  &  1.417  $\pm$  0.217  &  1.178  $\pm$  0.191  \\ 
    $\chi^2/N$ & 0.699 & 0.719 & 0.589 \\
\enddata
\end{deluxetable*}

\begin{deluxetable*}{cccc}
\tablecaption{Retrieved atmospheric parameters from Model 2a\label{tab:421_solutions}}
\tablewidth{0pt}
\tablehead{\nocolhead{} & \colhead{NEB} & \colhead{EZ} & \colhead{SEB (Outbreak)}}
\startdata
    $r_1$ &  5.0 $\mu m$ & 5.0 $\mu m$ & 0.75 $\mu m$ \\ 
    $P_1$ &  3.019  $\pm$  0.174  bar  &  3.353  $\pm$  0.276  bar &  4.817  $\pm$  0.622  bar \\ 
    $ \tau_1$ &  9.930  $\pm$  0.627  &  16.124  $\pm$  1.812  &  81.189  $\pm$  15.257  \\ 
    $P_{1T}$ &  0.122  $\pm$  0.025  bar  &  0.047  $\pm$  0.011  bar  &  0.147  $\pm$  0.029  bar  \\ 
    $r_2$ &  0.02 $\mu m$ & 0.2 $\mu m$ & 0.5 $\mu m$ \\ 
    $ \tau_2$ &  0.059  $\pm$  0.002  &  0.067  $\pm$  0.006  &  0.681  $\pm$  0.016  \\ 
    $r_3$ &  0.1 $\mu m$ & 0.15 $\mu m$ & 0.05 $\mu m$ \\ 
    $P_3$ &  0.010  $\pm$  0.002  bar  &  0.010  $\pm$  0.003  bar  &  0.010  $\pm$  0.003  bar  \\ 
    $ \tau_3$ &  0.009  $\pm$  0.002  &  0.010  $\pm$  0.002  &  0.009  $\pm$  0.002  \\ 
    $f$ &  0.822  $\pm$  0.130  &  0.781  $\pm$  0.125  &  1.265  $\pm$  0.188  \\
    $\chi^2/N$ & 0.763 & 0.607 & 0.678 \\
\enddata
\end{deluxetable*}

\begin{deluxetable*}{cccc}
\tablecaption{Retrieved atmospheric parameters from Model 2b\label{tab:422_solutions}}
\tablewidth{0pt}
\tablehead{\nocolhead{} & \colhead{NEB} & \colhead{EZ} & \colhead{SEB (Outbreak)}}
\startdata
    $r_1$ &  0.5 $\mu m$ & 2.5 $\mu m$ & 2.5 $\mu m$ \\ 
    $P_1$ &  3.011  $\pm$  0.381  bar  &  4.232  $\pm$  0.469  bar &  4.749  $\pm$  0.636  bar \\ 
    $ \tau_1$ &  50.89  $\pm$  9.233  &  39.856  $\pm$  6.742  &  48.432  $\pm$  9.668  \\ 
    $P_{1T}$ &  0.144  $\pm$  0.029  bar  &  0.061  $\pm$  0.013  bar  &  0.169  $\pm$  0.031  bar  \\ 
    $r_2$ &  0.02 $\mu m$ & 0.02 $\mu m$ & 0.02 $\mu m$ \\ 
    $ \tau_2$ &  0.149  $\pm$  0.008  &  0.013  $\pm$  0.001  &  0.056  $\pm$  0.004  \\ 
    $r_3$ &  0.1 $\mu m$ & 0.05 $\mu m$ & 0.15 $\mu m$ \\ 
    $P_3$ &  0.010  $\pm$  0.003  bar  &  0.010  $\pm$  0.003  bar  &  0.010  $\pm$  0.003  bar  \\ 
    $ \tau_3$ &  0.009  $\pm$  0.002  &  0.007  $\pm$  0.002  &  0.009  $\pm$  0.002  \\ 
    $f$ &  1.117  $\pm$  0.215  &  0.826  $\pm$  0.133  &  0.919  $\pm$  0.148  \\ 
    $\chi^2/N$ & 0.522 & 0.600 & 0.527 \\
\enddata
\end{deluxetable*}

\subsection{Constraining particle size and cloud base}

\begin{figure*}
\centering
\includegraphics[scale=0.6]{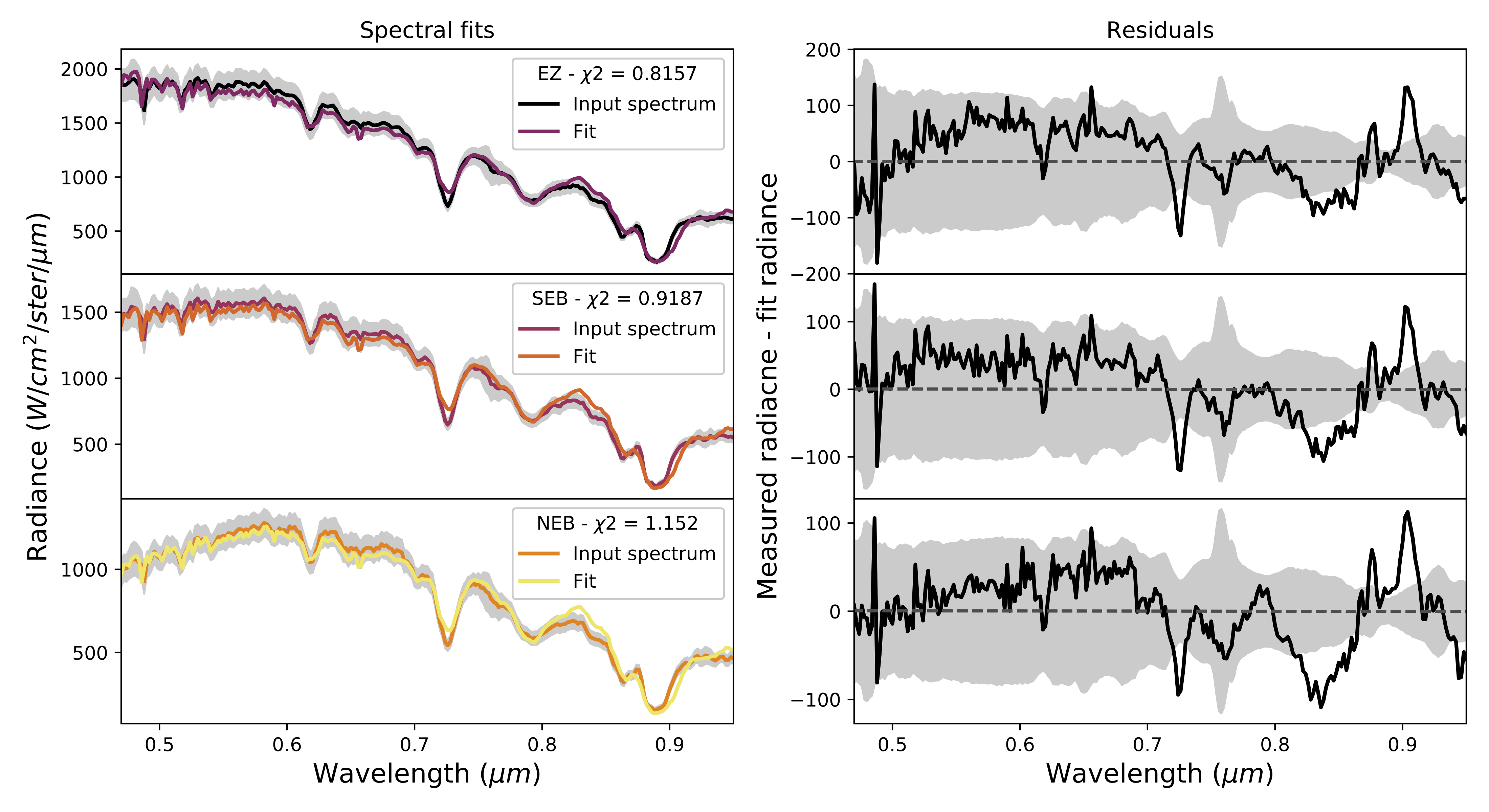}
\caption{Spectral fits resulting from testing more highly-constrained cloud models where we set our prior particle sizes to $r_1=1.0 \mu m$, $r_2=0.15 \mu m$, $r_3=0.1 \mu m$, and $P_1=3.0$ bars. For the rest of the input parameters, we used prior cloud structure values from \citet{sromovsky_2017} and allowed a 25\% variance on those parameters as with Models 1a-2b. Gray envelopes represent the combined observational and forward model uncertainty.}
\label{fig:constrained_spectral_Fits}
\end{figure*}

The results of Models 1a-2b are unfortunately rooted in the degeneracy between pairs of certain parameters, specifically main cloud base/optical depth and particle size/optical depth for a given layer. In order to better compare the cloud bands to one other, we ran an additional set of models where we held the particle sizes of each layer and $P_1$ fixed to reasonable values. This way, the optical depths of each layer and the top of the main cloud would be directly comparable to each other for each cloud band. We fixed $r_1$ to 1.0 $\mu m$ (a seemingly reasonable assumption based on the range of our best-fit particle sizes and the results from \citet{sromovsky_2017}), $r_2$ to 0.15 $\mu m$ (based on the assertion of \citet{baines_2019} that the chromophore particle should be in the 0.1-0.2 $\mu m$ range), and $r_3$ to 0.1 $\mu m$ (based on the assumptions of \citet{braude_2020} and \citet{sromovsky_2017}). We also held the cloud base constant at 3 bars for each of the cloud bands but used the same \textit{a priori} values and errors as we did in Models 1a-2b. We did also test shallower cloud bases at 1.0 and 1.4 bars (where 1.0 and 1.4 bars were in line with continuous aerosol profile retrieval results from \citet{braude_2020} for the EZ and NEB, respectively) but those were wholly unable to fit our spectra with our chosen fixed particle sizes and without allowing $n_2$ to vary. Again, due to the parameters' degeneracy it is possible that those pressures might fit our clouds with different particle sizes, or if we allow $n_2$ to vary in order to affect the overall absorptivity of large swaths of the spectrum. Regardless, the spectral fits from these highly constrained models and the retrieved parameters can help us understand the cloud bands relative to each other, and the results can be found in Figure \ref{fig:constrained_spectral_Fits} and Table \ref{tab:constrained_solutions}. 

\begin{deluxetable*}{cccc}
\tablecaption{Results of constrained atmospheric models\label{tab:constrained_solutions}}
\tablewidth{0pt}
\tablehead{\nocolhead{} & \colhead{NEB} & \colhead{EZ} & \colhead{SEB (Outbreak)}}
\startdata
    $r_1$ &  1.0 $\mu m$ & 1.0 $\mu m$ & 1.0 $\mu m$ \\ 
    $P_1$ &  3.0  bar  &  3.0   bar &  3.0  bar \\ 
    $ \tau_1$ &  21.326  $\pm$  0.7  &  42.825  $\pm$  1.401  &  34.204  $\pm$  1.397  \\ 
    $P_{1T}$ &  0.111  $\pm$  0.023  bar  &  0.048  $\pm$  0.011  bar  &  0.143  $\pm$  0.027  bar  \\ 
    $r_2$ &  0.15 $\mu m$ &  0.15 $\mu m$ & 0.15 $\mu m$ \\
    $ \tau_2$ &  0.256  $\pm$  0.007  &  0.037  $\pm$  0.005  &  0.134  $\pm$  0.007  \\ 
    $r_3$ &  0.1 $\mu m$ &  0.1 $\mu m$ & 0.1 $\mu m$ \\
    $P_3$ &  0.010  $\pm$  0.003  bar  &  0.010  $\pm$  0.003  bar  &  0.010  $\pm$  0.003  bar  \\ 
    $ \tau_3$ &  0.011  $\pm$  0.003  &  0.009  $\pm$  0.002  &  0.009  $\pm$  0.002  \\ 
    $f$ &  0.862  $\pm$  0.122  &  1.069  $\pm$  0.164  &  1.210  $\pm$  0.176  \\
    $\chi^2/N$ & 1.15 & 0.815 & 0.918 \\
\enddata
\tablecomments{These models held all cloud bases fixed at 3.0 bar and set the particle sizes for the main cloud, chromophore layer, and stratospheric haze to 1.0, 0.15, and 0.1 $\mu m$, respectively}
\end{deluxetable*}

After constraining particle sizes and cloud bases, the resulting chromophore, main cloud optical depths, and cloud-top pressures were significantly more in line with our predictions. The NEB has the most opaque chromophore layer, the lowest cloud top, and the least opaque main cloud of the three regions we measured. In contrast, the EZ has the least optically thick chromophore layer, the highest cloud top, and the optically thickest main cloud. The outbreak in the SEB, on the other hand, has characteristics that lie in between those we retrieved for the NEB and EZ (other than cloud top pressure, which is close to that of the NEB). This is indicative of the fact that this outbreak could have a morphology somewhere between the low, optically thin, red NEB and the upwelling, thick, bright white EZ. 

\subsection{Discrepancies in the spectral fits} \label{Discrepancies}

While our reduced $\chi^2$ values were low, there were common discrepancies in the fitted spectra for all models, as can be seen in the residuals of the spectral fits in Figures \ref{fig:all_residuals} and \ref{fig:constrained_spectral_Fits}. The spikes around 486-488 nm are the signature of a solar feature, potentially a hydrogen Balmer line, being one datapoint off from the location of this same feature in our data. Since this offset seems to only produce residuals with this particular feature and not any others, we have good reason to believe that it is a minor wavelength calibration issue that only affected this small region of the spectrum and did not seem to affect our results.

Other regions of the spectrum associated with residuals outside of our observational uncertainty, namely the spikes associated with the depth of the 727-nm methane band, the continuum region near 830 nm, and the width of the 890-nm methane band, might be addressed by changing some fundamental characteristics of the cloud structure in the models. In order to understand what might be causing the fitting issue in the 727-nm methane band specifically, we tested two modifications of the CB model. Since the methane bands in Jupiter's visible spectrum are highly sensitive to changes in vertical structure, we tested variations of the fractional scale heights (FSH) of our clouds and how the addition of a second cloud to the CB model might affect the methane bands.

In our models, we normally held the FSH of our clouds constant at a value of 1. It is possible, however, for a highly convective region (such as the EZ) to have a higher FSH. We tested our FSH assumption by first allowing a FSH of 1 to vary by 100\% (which resulted in a retrieved value of $\sim$1.5), and then by forcing it to stay close to 2. The higher fixed and retrieved FSH values did slightly deepen the 727-nm band, but they also dimmed the rest of the continuum, thereby obviating any possible improvement to the 727-nm band. Conducting similar tests with the SEB spectrum actually degraded the fit by a significant amount. Since the NEB is a region of downwelling, we did not test an FSH value of 2 but we did allow the FSH to vary by 100\%. The fit was not improved in this case, either.

We conducted a second alteration to our atmospheric models by adding an extra sheet cloud below the default CB cloud layers, as \citet{simon_2001b} found that some regions of the NEB and quiescent SEB were best fit by a model that included an extra sheet cloud. We found that the composition and particle size of the deep cloud made no difference in the output. The deep cloud was arbitrarily placed at a pressure of $\sim$5 bars with a top at $\sim$4 bars, given an optical depth of 5.6, and an arbitrary particle size of 0.4 microns. We found that this extra cloud did not improve the fit nor did it conserve the brightness of the continuum while simultaneously deepening the 727-nm band. We did find that this deep cloud  afforded minute changes to the spectral fit in ways that simply increasing the optical depth of the main cloud could not. However, the current iteration of NEMESIS is limited in its ability both to implement this deep cloud and simultaneously retrieve both its characteristics and those of the CB model cloud, so we were unable to further pursue and more rigorously test this modified version of the CB model. That being said, it is possible to alter the software within NEMESIS and implement such a model, although such a modification is outside the scope of this work.

While these changes to the CB model did not eliminate the discrepancies in our spectral fits, we cannot rule out some other modification that might better improve the fits. \citet{braude_2020}, after retrieving their own universal chromophore, retrieved a continuous combined cloud/haze profile that better fit their spectra than the CB model alone. Therefore, it is possible that there is some other cloud parameterization that, when combined with a universal chromophore, can provide better spectral fits that eliminate the fitting issues that we experienced. 

\section{Discussion and Summary} \label{summary}

In this study, we set out to accomplish two tasks: to test the validity of the CB parameterization of the structure of Jupiter’s uppermost cloud deck and to determine whether or not the \citet{carlson_2016} chromophore is the best suited chromophore candidate to use in the thin chromophore layer of the CB model. To do so, we conducted radiative transfer modeling of visible spectra extracted from three of Jupiter’s major banded regions: the NEB, the EZ, and from an outbreak region within the SEB. These data were pulled from hyperspectral image cubes obtained in March 2017 during \textit{Juno}’s 5$^{\mathrm{th}}$ perijove pass at the 3.5-m telescope at APO in Sunspot, NM. 

The results of our five sets of models show that the CB model provides a satisfactory fit, with reduced chi squared values below or on the order of 1, even when we assume various \textit{a priori} particle sizes for each of the cloud layers, when we did not allow the imaginary index of refraction of the chromophore to vary, and when we held cloud base and particle size constant between each cloud feature. In fact, adding more degrees of freedom to the models by allowing the imaginary index of the chromophore to vary did not noticeably improve our spectral fits. However, when we allowed $n_2$ to vary, we found that a different chromophore might provide much more physically realistic optical depths, especially in the SEB outbreak region. 

While the CB model consistently fit our spectra well, there were some persistent issues with fitting certain portions of the spectra, specifically the depth of the 727-nm methane band, the continuum level around 820 nm, and the width of the 890-nm methane band, that we were unable to mitigate completely by changing the main cloud FSH or by adding a deep cloud to the model. While our modifications to the CB model did not improve our spectral fits, we cannot rule out that some other change in parameterization can help, such as a continuous aerosol profile like those retrieved by \citet{braude_2020}. However, pinpointing the exact modification to the model or developing a replacement for the CB model is outside the scope of this work as we sought only to confirm whether or not the CB model was a valid parameterization of the troposphere at the locations and geometries we analyzed.

\begin{figure*}
\centering
\includegraphics[scale=0.7]{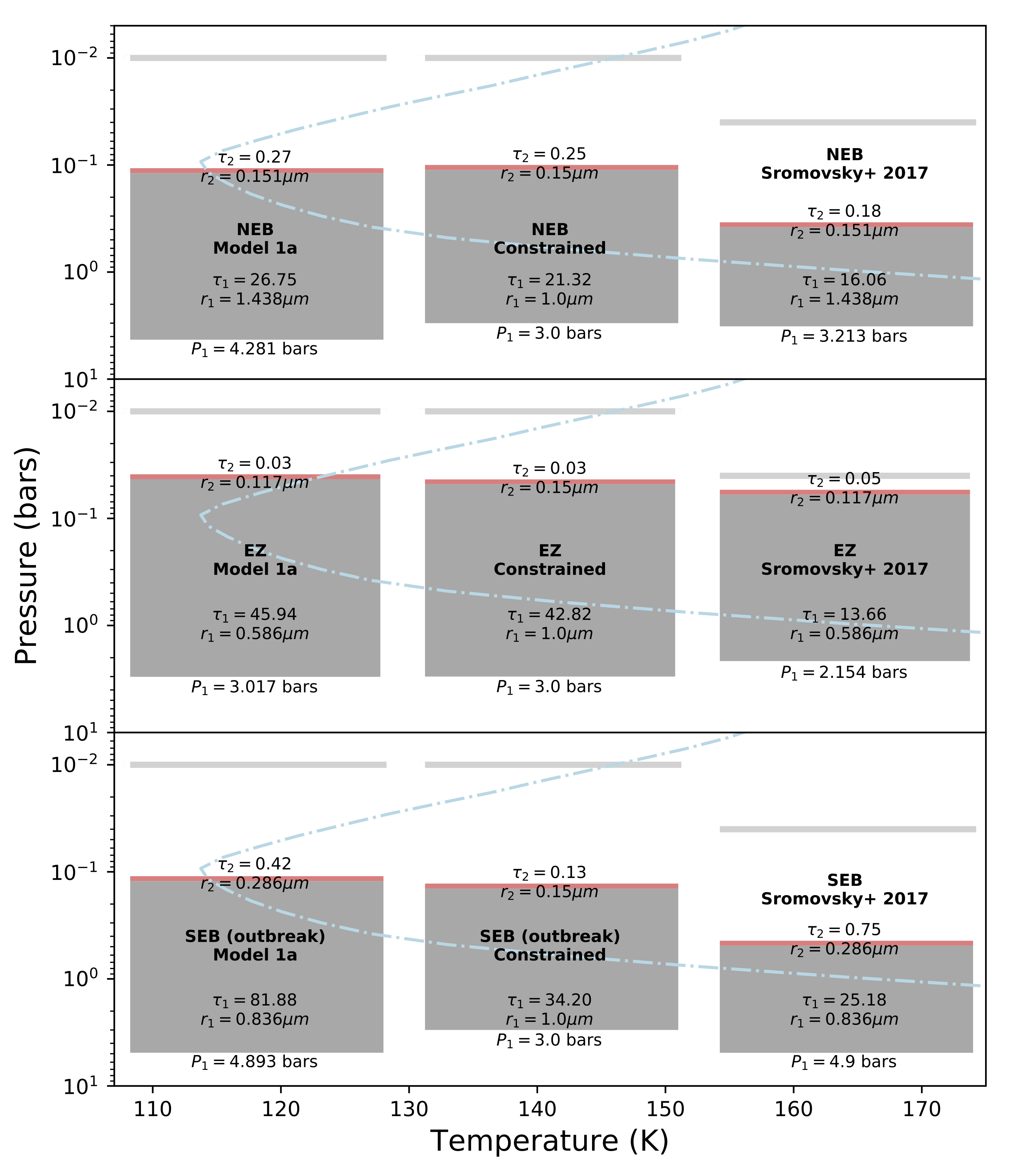}
\caption{Comparison of cloud structure results from Model 1a and our constrained models to the results of \citet{sromovsky_2017} for each cloud feature. Model 1a is the most directly comparable to the work of \citet{sromovsky_2017} since all of its inputs were from that work and we did not allow $n_2$ to vary. Dashed blue line represents the temperature profile used in this work.}
\label{fig:sromovsky_comparison_1}
\end{figure*}

\subsection{Cloud structure}

Our results for the cloud structure parameters (base pressure, top pressure, and optical depth) are most directly comparable to those of \citet{sromovsky_2017}, who did not deviate from the CB cloud structure and who also modeled the NEB, EZ, and a quiescent region of the SEB. In Figure \ref{fig:sromovsky_comparison_1}, we show comparisons of our cloud structure results from Models 1a (wherein all of our \textit{a priori} parameter values were from \citet{sromovsky_2017} and we did not let $n_2$ vary) and the constrained models where we held cloud base and particle size constant to the results from \citet{sromovsky_2017} when they assumed an ammonia-dominated cloud and used multiple viewing geometries to constrain their models.

The EZ clouds are the most similar in their physical extent between these two bodies of work, but the clouds that we retrieved in Model 1a were considerably more optically thick than the result in \citet{sromovsky_2017}. Our Model 1a EZ cloud does have a deeper base than the \citet{sromovsky_2017} EZ, which due to the positive correlation between optical depth and cloud base might produce a more optically thick cloud, but when we compare optical depth bar$^{-1}$ as a measure of whether or not we are comparing equally degenerate solutions, we find that of the EZ cloud was over twice as high as the results from \citet{sromovsky_2017} at 15.4 bar$^{-1}$ and 6.5 bar$^{-1}$, respectively. In other cases, major differences in optical depth could be explained by differences in particle size which result from the degeneracy between optical depth and particle size, but since the particle sizes between Model 1a and the EZ model from \citet{sromovsky_2017} are identical, it is possible that these results show just how much the EZ structure changed in the 17 years between observations. What is very consistent, however, is the optical depth of the chromophore layer; for each case, we see very optically thin chromophore layers over the bright EZ. 

Our retrieved NEB cloud from Model 1a is much more extended into the atmosphere than the one calculated by \citet{sromovsky_2017}, and we found higher optical depths for both the chromophore layer and the main cloud. Since we used the same particle sizes as \citet{sromovsky_2017} in Model 1a, these differences again can’t be explained by the degeneracy between particle size and optical depth. However, the difference between optical depths for these NEB retrievals is smaller than the difference in the EZ optical depths. Again, these differences could be indicative of how much Jupiter’s NEB band has changed in the 17 years since \textit{Cassini}’s flyby of the planet.

While we can compare the EZ and NEB regions between this work and \citet{sromovsky_2017}, the measurements of the SEB are not directly comparable. \citet{sromovsky_2017} measured a quiescent region of the SEB, while we tested a bright outbreak storm that was blooming within the SEB cloud deck. While these spectra were both extracted from approximately the same latitude, they probe very different cloud features, so they are not directly comparable as a way to understand how the quiescent SEB changes over time or as a way to check the degree to which our results are realistic. We can, however, use these results to understand how the structure of an outbreak cloud differs from that of a quiet SEB. In Model 1a and 2a, when we used different particle sizes but we did not allow $n_2$ to vary, we retrieved a much more optically thick ($\tau_1>80$) outbreak cloud than what \citet{sromovsky_2017} found for the quiescent SEB, but to the point where it could be considered unphysical. These results are consistent with the models of the 2009-2010 SEB fade from \citet{perez_hoyoz_2012_SEB}, which similarly found that the density of the SEB tropospheric sheet cloud became so great ($\tau >$ 100) that they were no longer sensitive to its properties. This consistency between results could either reaffirm that such outbreak or brightening events do result in incredibly dense cloud features, or that the cloud layering schema used in both this work and \citet{perez_hoyoz_2012_SEB} is not an accurate parameterization for such weather events. We also found that, while the satisfactory spectral fits to the other belt we measured were largely independent of the prior particle size values that we tested in Models 2a and 2b, the SEB outbreak was considerably more sensitive to input particle sizes and only produced well-fit spectra for less than half of those 108 particle size combinations.

The CB cloud layering scheme, under various \textit{a priori} assumptions and constraints, produced more physically realistic and more often well-fitting results for the NEB and EZ but not for the SEB outbreak. This points to the fact that the CB cloud layering scheme might not be an accurate approximation of the troposphere during significant weather events and that some other parameterization might be necessary in such cases.

\begin{figure*}
\centering
\includegraphics[scale=0.6]{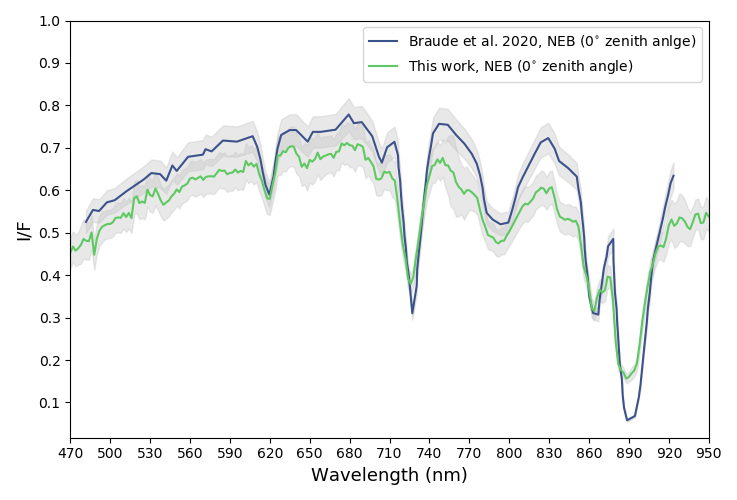}
\caption{A direct comparison of the 0$^{\circ}$ emission zenith angle NEB spectra from \citet{braude_2020} and this work. Differences between the spectra are likely a result of differences in photometric calibration, instrument and filter shape, and even the physical changes in the NEB. Gray envelopes indicated observational uncertainty; we assumed an average error of 5\% for the \citet{braude_2020} spectrum, which is approximately the value reported in their work.}
\label{fig:neb_comparison}
\end{figure*}

\citet{braude_2020} retrieved continuous cloud profiles for several cloud regions, including the EZ and NEB. For those cloud regions, the pressures where the optical depth of the aerosol profile was at its maximum value, which served as a proxy for cloud base pressure, of $1.4\pm0.1$ bars for the EZ and $1.07\pm0.08$ bars for the NEB. When we conducted our retrieval tests when $P_1$ was tightly constrained, we tested bases of 1.0 and 1.4 bars, but were entirely unable to fit our spectra even if we allowed all other parameters to vary. Figure \ref{fig:neb_comparison} shows a direct comparison of the 0$^{\circ}$ emission zenith angle NEB spectrum from this work and \citet{braude_2020}. The spectrum from \citet{braude_2020} is brighter at almost all wavelengths, especially along the continuum. Based on this relative difference, it is interesting that we were unable to retrieve optically thinner clouds in the NEB at the same approximate base pressure as \citet{braude_2020}. Our NEB cloud from Model 1a did have a similar optical depth bar$^{-1}$ value, but the rest of our models were either more or less optically thick as a function of pressure than those retrieved by \citet{braude_2020}. It is interesting to note that while the dates analyzed in \citet{braude_2020} bracket the perijove pass we analyzed here and despite using the same radiative transfer code, there are discrepancies between our results. This is likely a result of the overarching issue mentioned in Section \ref{sec:intro}, where simple and subtle discrepancies between datasets and methodology can be the determining factor in whether a given model fits one dataset but not another.

An important issue with these cloud structure results is that the retrieved cloud bases for these NH$_3$-dominated clouds, in this work and in both \citet{braude_2020} and \citet{sromovsky_2017} as the authors note, lie well below the level where the temperature becomes too high for ammonia ice to condense. A potential explanation is that we are actually approximating two or even three tropospheric clouds as one within the CB model. These approximated clouds could be composed of NH$_4$SH, H$_2$O, or a mixture of both. Our deep cloud tests revealed that we had difficulty resolving between a single main cloud and a CB model that included an extra, deeper sheet cloud. We also found that different compositions for this second cloud did not affect the output spectrum. Therefore, it is possible that we are simply not sensitive to such differences in cloud structure or to deeper clouds with different compositions. If this is the case and we are actually sensing different cloud layers and approximating them as one, this allows for the possibility that the deeper layers are of some other composition or structure.

However, in the case of the SEB outbreak, it is possible that these deep cloud bases are in fact the bases of water clouds. It has been theorized that outbreaks begin at the water cloud level ($\sim$6 bars) and, driven by strong moist convection, bloom up through the other cloud layers \citep{depater_2019}. Other ground-based measurements of this same outbreak in the SEB (but in the infrared regime) showed preliminary evidence of clouds with bases near the 5-bar level, supporting this theory that SEB outbreaks begin at the water-cloud level \citep{bjoraker_2018}. Observations and radiative transfer of this particular outbreak with the Atacama Large Millimeter/Submillimeter Array (ALMA) also showed that it was ``consistent with models where energetic plumes are triggered via moist convection at the base of the water cloud" \citep{depater_2019}. Therefore, with these results from other wavelength regimes, it is probable that the cloud bases we retrieved in Models 1a through 2b for the SEB outbreak region are accurate and are further evidence for the cloud’s origins in the deep water cloud.

\begin{figure*}
\centering
\includegraphics[scale=0.6]{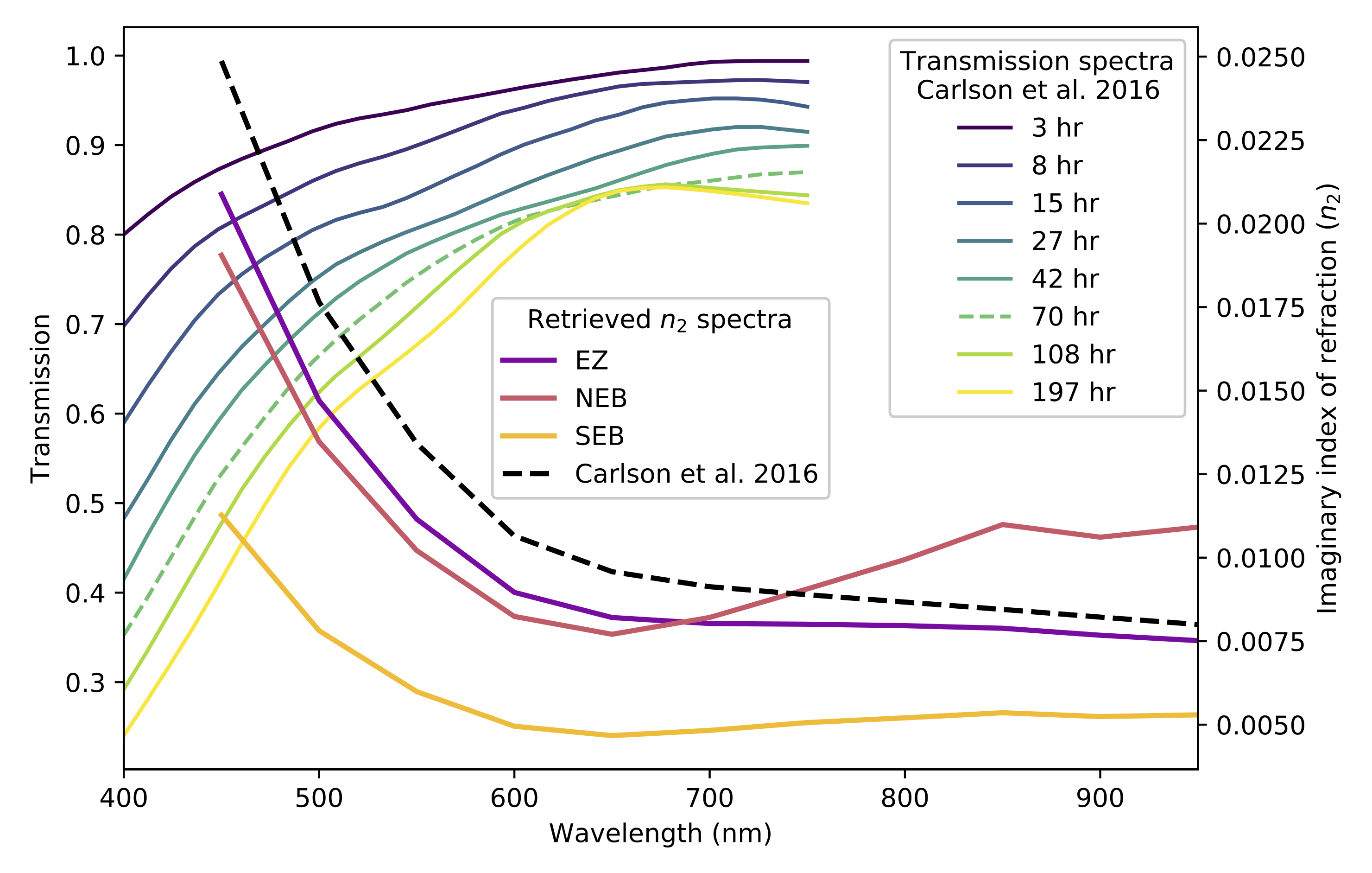}
\caption{A comparison of the transmission spectra of the \citet{carlson_2016} chromophore at different points in time during the process of irradiation as reported in \citet{carlson_2016} to imaginary indices of refraction retrieved in Model 1b. The 70-hour transmission spectrum (green dashed line) was used to derive the imaginary index of refraction (black dashed line) \citep{carlson_2016}. While it is nontrivial to derive comparable $n_2$ spectra from each transmission curve, we can see a pattern of brightening and the development of a shallower blue slope as we move from ``older" to ``younger" chromophores. In the imaginary indices of refraction, we see that the EZ and the SEB outbreak produced less-absorbent chromophores at all wavelengths, potentially showing signs of such a ``younger" chromophore.}
\label{fig:chromo_comparison_2}
\end{figure*}

\begin{figure*}
\centering
\includegraphics[scale=0.6]{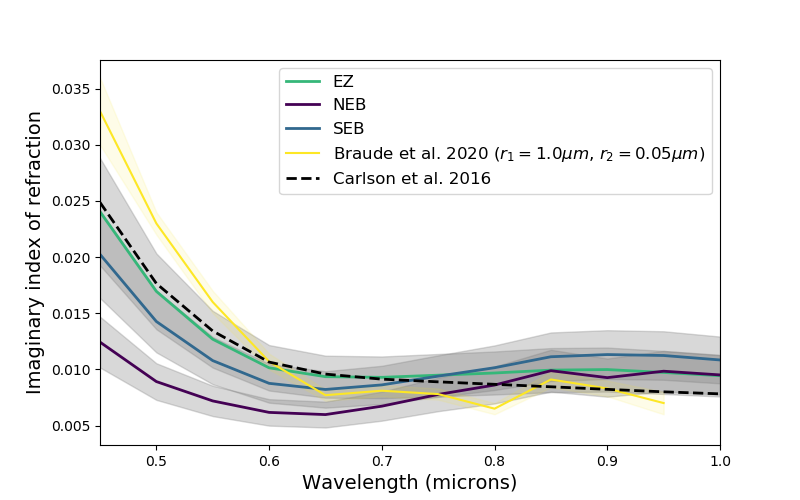}
\caption{Our chromophore retrieval results from Model 2b alongside results from \citet{braude_2020} when $r_1=1.0~\mu m$ and $r_2=0.05~\mu m$ were used.}
\label{fig:chromo_comparison}
\end{figure*}

\subsection{Chromophore imaginary index}

We noted in the previous subsection that the CB model produces unrealistically thick clouds within the SEB outbreak regardless of input particle size and if we do not allow the imaginary index of the chromophore to vary, which is evidence for a change in the CB model being necessary for such major weather events. However, when we assumed particle sizes from \citet{sromovsky_2017} and allowed the imaginary index to vary in Model 1b, we found that the results for the EZ and the SEB outbreak, both of which are bright, upwelling regions of Jupiter’s troposphere, echoed transmission functions of less-irradiated samples of the \citet{carlson_2016} chromophore. While \citet{carlson_2016} published the imaginary index of refraction spectrum of their NH$_3$-based chromophore as derived from the transmission curve of the sample after it had been irradiated for 70 hours, they also presented the transmission curves from ``younger” samples, down to 3 hours. See Figure \ref{fig:chromo_comparison_2} for the transmission curves of the \citet{carlson_2016} chromophore irradiated for different amounts of time compared to the 70-hour imaginary index of refraction spectrum and the retrieved $n_2$ spectra for Model 1b. These ``younger" transmission curves share a similar shape to our retrieved $n_2$ spectra, as they are  generally less absorbent across all wavelengths than the 70-hour chromophore and the blue-absorbent slope tends to become shallower for ``younger" chromophores. 

If we were to take similar changes in absorbency (overall brightness and changes in blue slope) of an imaginary index of refraction spectrum as indicators of a younger chromophore, our $n_2$ retrieval results for the EZ and SEB outbreak in model 1b show evidence for such a chromophore that has been less exposed to solar irradiation.  The highly convective nature of these regions could be the source of this younger version of the \citet{carlson_2016} chromophore. By constantly overturning the fresh material at the cloud tops, there is not enough time for the attendant ammonia and acetylene gases to be processed into the 70-hour chromophore. Even if it does get processed to that age, it could be covered up by new cloud material. 

When we tested a wide range of particle sizes, our retrieved $n_2$ spectrum for Model 2b did not show the same evidence for chromophore age difference. We therefore compare these results to those of \citet{braude_2020}, who did not assume particle sizes from \citet{sromovsky_2017} and retrieved a new universal chromophore using NEB spectra. Interestingly, the parts of our $n_2$ retrievals for the NEB that were more or less absorbent than the \citet{carlson_2016} chromophore were almost exactly the opposite of what \citet{braude_2020} found for comparable particle sizes in the NEB. See Figure \ref{fig:chromo_comparison} for a direct comparison of our $n_2$ retrieval from Model 2b and the $n_2$ retrieval from \citet{braude_2020} when they used $r_1=1.0~\mu m$ and $r_2=0.05~\mu m$. To quantify these differences between these complex index of refraction spectra, we took the ratio of a given imaginary index of refraction at 500 nm to its value at 650 nm. The resulting ratios were 1.842$\pm$0.130 for \citet{carlson_2016} (assuming 5\% error), 2.987$\pm$0.233 for the result from \citet{braude_2020} shown in Figure \ref{fig:chromo_comparison}, and 1.813$\pm$0.414, 1.492$\pm$0.346, and 1.737$\pm$0.396 for the EZ, NEB, and SEB outbreak $n_2$ results respectively from Model 2b. Qualitatively, this means our retrieved $n_2$ spectrum for the NEB was less absorbent than the \citet{carlson_2016} chromophore at wavelengths shorter than $\sim$650 nm but also more absorbent past $\sim$800 nm, while at similar particle sizes to the best-fit sizes in Model 2b \citet{braude_2020} found a chromophore that was more absorbent than \citet{carlson_2016} at wavelengths shorter than $\sim$650 nm and less absorbent at longer wavelengths (except at 850 and 900 nm).This difference might point either to changes in the NEB between observations analyzed here and in \citet{braude_2020}, or to simple discrepancies between our datasets that might have propagated through to affect our results in such a way.

\subsection{Juno ammonia measurements}

Analysis of the microwave measurements from the MicroWave Radiometer (MWR) on board the \textit{Juno} spacecraft have shown that the deep ammonia gas abundance (below the 0.7-bar level) in Jupiter's atmosphere varies strongly as a function of latitude: there is an enhanced column of ammonia gas below the EZ but depletions at the mid-latitudes, especially in the NEB \citep{guillot_2019, li_2017}. These relative abundances appear to be consistent across many different perijove passes. To compare the ammonia profile scaling factors we retrieved in this work to the ammonia abundances as observed by the MWR instrument on board \textit{Juno}, we utilized ammonia retrievals from Cheng Li (pers. comm.) from PJ5 MWR measurements. To find a comparable scaling factor, we divided the \textit{Juno}-measured VMR value at the 1-bar level by the deep VMR that we assumed in this work (0.0002). We calculated this \textit{Juno}-observed scaling factor at the latitudes we observed for the EZ and NEB ($\sim-3^\circ$ and 13$^\circ$ W, respectively). 

The analogous MWR-measured PJ5 scaling factors were 1.24 for the EZ and 0.752 for the NEB. The models for which we measured relative enhancements and depletions of ammonia in the EZ and NEB respectively were in Model 1a and Model 1b. However, we saw a reversal of that relationship when we tested the particle size grids in Models 2a and 2b when the NEB was relatively enhanced in ammonia gas while the EZ was depleted. The fact that our ammonia scaling factor results in Models 1a and 1b reflect the relative enhancements and depletions as observed by \textit{Juno} could be a sign that the prior assumptions we made in those models are closer to reality: namely, the particle sizes calculated by \citet{sromovsky_2017} were correct, or at least closer to the real particle sizes than the final ones we used from our size grids in Models 2a and 2b. Future analysis will require a more detailed retrieval of the ammonia gas profile beyond a simple scaling factor in order to produce results that are directly comparable to those from \textit{Juno}.

\subsection{Summary}

In this work, we tested the validity of the Cr\`{e}me Br\^ul\'ee (CB) model of Jupiter’s uppermost cloud deck by both testing the cloud layering scheme of the model and by measuring the degree to which the \citet{carlson_2016} chromophore is truly universal within this parameterization.  We found that the CB cloud structure parameterization and \citet{carlson_2016} chromophore provided sufficient fits to our data with reduced $\chi^2$ values below or on the order of 1 even when we used different prior assumptions, with the exception of the SEB outbreak, which was far more sensitive to particle size assumptions than the NEB and EZ. Allowing the complex index of refraction of the chromophore layer to vary did not significantly improve our spectral fits, despite adding more degrees of freedom to the parameter space.  

While we found deep cloud bases for almost each set of prior assumptions, the base of the SEB outbreak region was often much deeper relative to the base of the cloud in the NEB and EZ. This could be evidence of a deep cloud base that might have its origins at the base of the water cloud, confirming similar observations made in different wavelength regimes and furthering the theory that convective outbreaks begin in the deep water cloud. Both of the highly convective regions that we measured (the SEB outbreak and the EZ) showed evidence of a younger version of the \citet{carlson_2016} chromophore. This points to the fact that the overturning material at the tops of these clouds keep the \citet{carlson_2016} chromophore from being irradiated less than 70-hour \citet{carlson_2016} chromophore. 

It should be noted that both the limited viewing geometries of the analyzed spectra and the degeneracy (or near-degeneracy) between multiple parameters in the visible wavelength regime, namely between optical depth/cloud base and optical depth/particle size, limited our ability to definitively retrieve certain parameters, and therefore limited our ability to come to concrete conclusions about the nature of Jupiter’s clouds. Future modeling of Jupiter’s visible spectrum would benefit both from greater statistical analyses, such as the use of Markov chain Monte Carlo methods to test a much larger population of parameter combinations, and from the combined use of visible and other adjacent wavelength regimes at multiple viewing geometries in order to better constrain these parameter spaces between different datasets.

\acknowledgements

The author gratefully acknowledges Ashwin Braude, Rohini Giles, Alexander Thelen, and the rest of Team NEMESIS for their invaluable advice regarding the use of NEMESIS and the modeling of Jupiter's complex atmosphere; David Kuehn, Robert Hull, Hanyu Zhan, and Hannah Gallamore for their contributions to NAIC observations, software, and lab measurements; Mike Wong and Larry Sromovsky for helpful discussions regarding various aspects of this project; and the anonymous reviewers whose helpful comments and suggestions greatly improved the quality of this manuscript. This work was supported by NASA’s Minority University Research and Education Project (MUREP) NASA Fellowship Activity through training grant number 80NSSC18K1701 and by Research Support Agreement 1569980 from the Jet Propulsion Laboratory, as a subaward of a NASA/Solar System Observations grant. Glenn Orton, Kevin Baines, and James Sinclair were supported by funds from NASA distributed to the Jet Propulsion Laboratory, California Institute of Technology. A portion of this research was carried out at the Jet Propulsion Laboratory, California Institute of Technology, under a contract with the National Aeronautics and Space Administration.

\bibliography{main}{}
\bibliographystyle{aasjournal}

\end{document}